\documentclass{emulateapj}

\def\OII{[{\ion{O}{2}}]}

\def\NIIHa{[{\ion{N}{2}}]/H$\alpha$}
\def\SIIHa{[{\ion{S}{2}}]/H$\alpha$}

\def\OIIIHb{[{\ion{O}{3}}]/H$\beta$}
\def\OIII5007Hb{[{\ion{O}{3}}] $\lambda5007$/H$\beta$}
\def\4959_5007{[\ion{O}{3}] $\lambda \lambda$4959,5007}
\def\OIII49595007{[\ion{O}{3}] $\lambda \lambda 4959,5007$}
\def\ratioR23{([\ion{O}{2}] $\lambda$3727 +[\ion{O}{3}] $\lambda\lambda$4959,5007)/H$\beta$}
\def\R23{${\rm R}_{23}$}
\def\dS23{${\rm S}_{23}$}

\def\Msun{${\rm M}_{\odot}$}

\def\NII{[{\ion{N}{2}}]}

\def\NIIOII{[\ion{N}{2}]/[\ion{O}{2}]}

\def\OH{$\log({\rm O/H})+12$}

\def\ratioS23{([\ion{S}{2}] $\lambda \lambda$6717,31 +
[\ion{S}{3}] $\lambda\lambda$9069,9532)/H$\beta$}
\def\NIIHa{[\ion{N}{2}]/H$\alpha$}

\def\Hb{{H$\beta$}}
\def\O4363{[{\ion{O}{3}}] $\lambda$4363}
\def\OIII{[{\ion{O}{3}}]}

\def\Ha{{H$\alpha$}}
\def\L60{L$_{60}$}

\def\HII{\ion{H}{2}}
\def\B26{${{\rm B}_{26}}$}
\def\I26{${\rm I_{B_{26}}}$}
\def\Inuc{${\rm I_{nuc}}$}
\def\Iint{${\rm I_{int}}$}
\def\tsim{$\sim$}

\shorttitle{}
\shortauthors{}

\begin{document}

\title{Aperture effects on Star Formation Rate, Metallicity and Reddening}

\author{Lisa J. Kewley\altaffilmark{1}}
\affil {Harvard-Smithsonian Center for Astrophysics}
\affil{60 Garden Street MS-20, Cambridge, MA 02138}
\email {lkewley@cfa.harvard.edu}
\altaffiltext{1}{CfA Fellow}

\author{Rolf A. Jansen}
\affil{Dept. of Physics \& Astronomy, Arizona State University}
\affil{P.O. Box 871504, Tempe, AZ 85287-1504}

\author{Margaret J. Geller}
\affil{Smithsonian Astrophysical Observatory}
\affil{ 60 Garden Street MS-20, Cambridge, MA 02138}

\begin{abstract}
We use 101 galaxies selected from the Nearby Field Galaxy Survey (NFGS) to 
investigate the effect of aperture size on the star formation rate,
metallicity and reddening determinations for galaxies.  
Our sample includes galaxies of all Hubble types except ellipticals 
with global SFRs ranging from 0.01 to 100 
\Msun yr$^{-1}$, metallicities between $7.9\lesssim$\OH$\lesssim9.0$, and reddening 
between $0\lesssim A(V)\lesssim3.3$.  We compare the star formation rate,
metallicity and reddening derived from nuclear spectra to those derived
from integrated spectra.   For apertures capturing $<20$\% of the $B_{26}$
light, the differences between nuclear and global metallicity, extinction and
star formation rate are substantial.  Late-type spiral galaxies show
the largest systematic difference of $\sim0.14$ dex 
in the sense that nuclear metallicities are greater than the global 
metallicities.  Sdm, Im, and Peculiar types have 
the largest scatter in nuclear/integrated metallicities, indicating a
large range in metallicity gradients for these galaxy types or clumpy 
metallicity distributions.  
We find little evidence for systematic differences between nuclear 
and global extinction estimates for any galaxy type.  However, 
there is significant scatter between the nuclear and integrated 
extinction estimates for nuclear apertures containing $<20$\% of the $B_{26}$
flux.  
We calculate an `expected' star formation rate using our nuclear spectra 
and apply the commonly-used aperture correction method.  The expected
star formation rate overestimates the global value for early type
spirals, with large scatter for all Hubble types, particularly late types. The differences
between the expected and global star formation rates probably result from the assumption
that the distributions of the emission-line gas and the continuum are 
identical.  The largest scatter (error) in the estimated SFR occurs 
when the aperture captures $<20$\% of the  $B_{26}$ emission. 
We discuss the implications of these results for 
metallicity-luminosity relations 
and star-formation history studies based on fiber spectra.
To reduce systematic and random errors from aperture effects, we recommend
selecting samples with fibers that capture
$>20$\% of the galaxy light.  For the Sloan Digital Sky Survey and the 
2dFGRS, redshifts $z>0.04$ and $z>0.06$ are required, respectively, to ensure a covering
fraction $>20$\% for galaxy sizes similar to the average size, type, and 
luminosity observed in our sample.  Higher-luminosity samples and samples
containing many late-type galaxies require a larger minimum redshift to 
ensure that $>20$\% of the galaxy light is enclosed by the fiber.
\end{abstract}


\section{Introduction}
Unprecedented insight into galaxy evolution has recently been obtained with
very large spectroscopic surveys, including the Sloan Digital Sky 
Survey (SDSS) and the  2 degree Field Galaxy Redshift Survey (2dFGRS).  These spectroscopic 
surveys allow fundamental physical properties of galaxies 
to be studied with increasingly large datasets.  Galaxy star formation 
rates and the metallicity-luminosity relation 
can now be analysed as a function of redshift for samples containing
thousands of galaxies 
\citep[e.g.,][]{Baldry02,Nakamura03,Schulte04,Lamareille04,Tremonti04}.

Unfortunately, comparisons of the spectroscopic 
properties of nearby galaxies 
with those at higher redshifts can be severely hindered by aperture effects. 
The combination of a fixed-size aperture 
and radial gradients and variations in metallicity, star-formation, 
and extinction can lead to a bias in these properties that may mimic or
hide evolution as a function of redshift.  
Typical fiber diameters are $3^{\prime \prime}$ for 
the SDSS and $2^{\prime \prime}$ for the 2dFGRS.  Spectra of nearby galaxies taken 
through such small apertures cover only a small portion of the 
nearest galaxies and capture a small fraction of their integrated light.  

Early studies of the effect of aperture size on apparent galaxy properties 
were carried out decades ago.  \citet{deVaucouleurs61} and 
\citet{Hodge63} used various
aperture sizes to show that early-type galaxies are redder in their 
galaxy centers.
\citet{Tinsley71} pointed out that the color-aperture relation seen in 
early-type spiral galaxies leads to the observation of bluer colors at higher-z in
an aperture of small fixed angular size.

Recent investigations into aperture effects have focused on spectral 
classification and star formation rates.  
\citet{Zaritsky95} showed that galaxies in the 
Las Campanas Redshift Survey (LCRS) at redshifts $z<0.05$ may be
misclassified because the $3.5^{\prime \prime}$ fibers capture only the 
central fraction of the galaxies' light.  

The \Ha\ emission-line is commonly used to estimate star-formation rates
in galaxies with redshifts $z<0.3$.  \citet{Perez03} analysed the 
\Ha\ emission in a sample of local star-forming 
galaxies selected from the Universidad Complutense de Madrid (UCM) Survey.
P\'{e}rez-Gonz\'{a}lez et al. concluded that long-slit or fiber 
spectra of nearby galaxies can miss substantial fractions of the 
emission-line flux (typically 50-70\%).  They suggest supplementing 
emission-line fluxes with direct emission-line imaging to overcome this
bias.  However, for large samples covering a range of redshifts, 
\Ha\ imaging of every galaxy is not currently viable.

Flux calibration of spectra in fiber-based surveys is difficult.  Thus
\Ha\ emission-line equivalent widths (EWs) are often used along with an
$r-$band magnitude to estimate star formation rates. 
\citet{Gomez03} pointed out that aperture bias may result in a 
systematic increase in the observed \Ha\ equivalent width (EW) 
for higher redshift galaxies relative to lower redshift galaxies 
in the SDSS.  Gomez et al. analysed SDSS galaxies within three equal 
volume redshift bins between $0.05<z<0.095$.  They found no evidence 
for an increase in the median and 75th percentile of the \Ha\ EW 
distribution among these three bins. 

\citet{Hopkins03} investigated aperture effects on \Ha-based 
star-formation rates in the SDSS by comparing the \Ha\ SFRs with those 
derived from the radio.  They found that the \Ha\ SFRs are overestimated
for galaxies requiring the largest aperture corrections.  The galaxies 
requiring the largest aperture corrections are either the closest or the most
massive galaxies.  Because of the large scatter between the two SFRs, 
Hopkins et al. were unable to quantify this aperture effect.  

The effect of aperture on metallicity, extinction, and star formation rate 
has been difficult to quantify  without high quality nuclear and 
global spectra for galaxies spanning a broad range in Hubble type.
In this paper, we use the Nearby Field Galaxy Survey 
(NFGS; Jansen et al.\ 2000a,b) to investigate
aperture effects on star-formation rate, metallicity, and extinction.
The NFGS is ideal for an aperture effect investigation because it 
provides integrated and nuclear spectra for 198 objectively selected 
nearby galaxies.  The NFGS (described in Section~2)
spans the full range in Hubble type and 
absolute magnitude in the CfA redshift catalog, thus enabling an
investigation into aperture effects as a function of both galaxy type and 
luminosity.  In Section~3, we investigate and quantify the flux
coverage of our nuclear and integrated spectral apertures.  
In Section~4, we investigate the effect of
aperture on metallicity estimates.  We show the effect of aperture on 
extinction estimates in Section~5, and we study
the effect of aperture on star-formation rates in Section~6.  In 
each section we discuss the implications of our results in
terms of the SDSS and 2dFGRS fiber-based sky surveys and their current 
aperture-correction methods.  In Section~8, we conclude 
that a minimum flux covering fraction of $\gtrsim20$\% is required for reliable 
estimates of extinction, metallicity and star formation rate, for samples 
with a similar mean galaxy size to that of the NFGS.  
Throughout this paper, we adopt the flat 
$\Lambda$-dominated cosmology as measured by the WMAP experiment 
($h=0.72$, $\Omega_{m}=0.29$; Spergel et al. 2003).

\section{Sample Selection and Spectrophotometry \label{Sample}}

The NFGS is ideal for investigating aperture effects in galaxies because it
is an objectively selected (unbiased) sample with both integrated and nuclear
spectra.  \citet{Jansen00a} selected the NFGS sample by sorting the 
CfA1 redshift survey \citep{Davis83,Huchra83} into 1 mag-wide bins of 
$M_{Z}$.  Within each bin, the sample was sorted according to
CfA1 morphological type.  To avoid a strict diameter limit, which 
might introduce a bias against the inclusion of low surface brightness 
galaxies in the sample, Jansen et al. imposed a
radial velocity limit, ${\rm V_{LG} (km\,s^{-1}) > 10^{-0.19-0.2M_{z}}}$
(with respect to the Local Group standard of rest).  Galaxies in the
direction of the Virgo Cluster were excluded to avoid a sampling
bias favoring a cluster population.  Lastly, every  $Nth$ galaxy
in each bin was selected to approximate the local galaxy 
luminosity function \citep[e.g.,][]{Marzke94}.  
The final 198-galaxy sample 
contains the full range in Hubble type and absolute magnitude present 
in the CfA1 galaxy survey and is a fair representation of the local galaxy
population.   

\citet{Jansen00b} provide integrated and nuclear spectrophotometry 
for almost all galaxies in the NFGS sample.  They obtained 
integrated spectra by scanning a $3''$ slit across each galaxy. The resulting 
integrated spectra include 52-97\% of the light enclosed 
within the 
$B_{26}$ isophote, with an average $B_{26}$ fraction of 
82$\pm$7\% for the 198 galaxy NFGS sample.  
These spectra represent the luminosity weighted mean of the emission from
stars and \HII\ regions over a wide range in distance from the galaxy center.The integrated emission-line properties are therefore
weighted towards the highest surface brightness, largest, and 
least extincted \HII\ regions. 

The nuclear spectra were obtained with a $3^{\prime \prime}$ slit centered on the 
nucleus and aligned along the major axis of each galaxy, sampling
$3''\times 6 \farcs84$.  The covering
fraction of the nuclear spectra depends on the radial light profile and 
ranges between 0.4-72\% of the light within the 
$B_{26}$ isophote, with an average covering fraction of 10$\pm11$\%.

We have calibrated the integrated and nuclear fluxes to absolute fluxes 
by careful comparison with B-band surface photometry 
\citep[described in ][; hereafter Paper I]{Kewley02a}.  
We corrected the \Ha\ and \Hb\ emission-line fluxes 
for underlying stellar absorption as described in Paper I.  

We used two methods to correct the NFGS emission-line fluxes 
for Galactic extinction, based on: 
(1) the HI maps of \citet{Burnstein84}, as listed in the Third Reference Catalogue of Bright Galaxies \citep{deVaucouleurs91}, 
and (2) the COBE and IRAS maps (plus the Leiden-Dwingeloo maps 
of HI emission) of \citet{Schlegel98}.  The average Galactic extinction is
E($B-V$)=$0.014 \pm 0.003$ (method 1) or E($B-V$)=$0.016 \pm 0.003$ (method 2).

To reject galaxies with significant emission from an AGN, we used the nuclear and integrated optical emission-line ratios to classify the galaxies.  
For those galaxies with  \OIIIHb, \NIIHa, and (if available) \SIIHa\ ratios, we used the 
theoretical optical classification scheme developed by \citet{Kewley01a}.  In this 
scheme, galaxies that lie above the theoretical ``extreme starburst line" in the standard 
optical diagnostic diagrams are classed as 
AGN, while those that lie below the line are classed as HII region-like.  Galaxies that lie 
in the AGN region in one diagnostic diagram but in the HII region section of the other 
diagram are classed as ``ambiguous" (AMB) galaxies.  A fraction of galaxies in the NFGS (35/198)
have \NIIHa\ ratios but immeasurable \OIII\ or \Hb\ fluxes in their nuclear (24/198) and/or integrated (15/198) spectra.  These galaxies are classified as HII region-like 
if  log(\NIIHa)$<-0.3$, typical of starburst galaxies and HII regions \citep[e.g.,][]{Kewley01b,Dopita00}.  We are unable to classify those galaxies with $-0.3\leq $log(\NIIHa)$\leq 0.0$ if \OIIIHb\ is unavailable because such line ratios can be produced by both AGN and starburst galaxies.  Galaxies without \OIIIHb\ but with strong log(\NIIHa)$>0.0$ are classed as AGN \citep[e.g., Figure~1 of ][]{Brinchmann04}.
Our final adopted classification is based on the nuclear class if available and the integrated class 
otherwise.   We provide the global, nuclear, and adopted classes for the NFGS sample in Table~\ref{classtable}.  We give quality flags, $Q$,  for the global and nuclear classes based on 
the number of line ratios used for classification.  Quality flags $Q=$1, 2, and 3 correspond to the use of 
\{\OIIIHb, \NIIHa, \SIIHa\}, \{\OIIIHb, \NIIHa\}, and \NIIHa\ respectively.
Following this scheme, the emission-line ratios indicate that the global spectra of
118/198 NFGS galaxies are dominated by star formation, 12/198 are dominated by AGN, and 
 8/198 are ``ambiguous'' galaxies.   The remaining galaxies do not have sufficient emission-lines in their spectra to allow classification.   Many of these unclassified galaxies are ellipticals.   Because ambiguous galaxies are likely to contain both starburst and AGN  activity \citep[see e.g.,][]{Kewley01b,Hill99}, we do not include them in the following analysis.  

We note that the fraction of AGN found in surveys may depend on the aperture size.  Larger apertures may include more emission from extra-nuclear HII regions, changing the classification from AGN to HII region-like \citep{Storchi91}.  The NFGS does not
contain a sufficiently large number of galaxies with significant AGN
activity to investigate the effect of aperture on classification quantitatively.   Of the 
115 NFGS galaxies with both nuclear and global classifications, 2 galaxies change class from their integrated to nuclear spectra; A0857+5242 changes from HII region-like to an AGN class, and NGC~3165 changes from HII region-like to ambiguous.   It is not possible to investigate a variation in nuclear to integrated \NIIHa, \OIIIHb, or \SIIHa\ ratios in terms of a potential AGN contribution
because variations in these ratios are likely to be driven by gradients in metallicity (\NIIHa, \OIIIHb) and/or ionization parameter (\NIIHa, \OIIIHb, \SIIHa) in HII region-like galaxies \citep[e.g.,][]{Kewley01a}.

To calculate the extinction and star formation rates, 
we require measurable \Ha\ and \Hb\ emission-line fluxes. 
We calculate metallicities
using the \citet{Kewley02b} ``recommended'' method.  To utilize this
method, we also require \OII $\lambda3727$ and \NII $\lambda6584$ fluxes for 
log(\NIIOII)$\gtrsim -1.0$.  For log(\NIIOII)$\lesssim -1.0$, 
we require measureable \OII $\lambda3727$ and 
\OIII $\lambda \lambda 4959,5007$. 
A total of 101/118 HII region-like galaxies in the NFGS satisfy these criteria.  
These 101 galaxies
(Table \ref{sample_table}) constitute the sample we analyse here. 

Because of  low S/N ratios in the \OIII $\lambda4959$ emission-line, we used
the theoretical ratio \OIII $\lambda5007$/\OIII $\lambda4959 \sim 3$
to calculate the \OIII $\lambda4959$ flux.  Errors in the absolute 
metallicities are $\sim0.1$ in log(O/H)+12 units 
\citep[see][for a discussion]{Kewley02b}.  These errors primarily 
reflect the accuracy
of the theoretical models used to create the metallicity diagnostics.
Any error introduced by the diagnostics is likely to be systematic 
\citep[e.g.,][]{Kobulnicky04}.
Because we are using {\it the same} metallicity diagnostic for both nuclear 
and integrated spectra, errors in the {\it relative} metallicities should 
be $\ll 0.1$ dex.

We corrected the emission line fluxes for reddening using the Balmer 
decrement and the \citet{Cardelli89} 
(CCM) reddening curve.  We assumed an ${\rm R_{V}=A_{V}/{\rm E}(B\!-\!V)} = 3.1$ and an 
intrinsic H$\alpha$/H$\beta$ ratio of 2.85 (the Balmer decrement for case B 
recombination at T$=10^4$K and $n_{e} \sim 10^2 - 10^4 {\rm cm}^{-3}$;
Osterbrock~1989).  After removing underlying Balmer absorption,
6/101 galaxies have  ${\rm H}\alpha/{\rm H}\beta <2.85$ for both integrated
and nuclear spectra, 11/101 galaxies have 
${\rm H}\alpha/{\rm H}\beta<2.85$ 
for nuclear spectra and ${\rm H}\alpha/{\rm H}\beta>2.85$ 
for integrated spectra, and 3/101 galaxies have  ${\rm H}\alpha/{\rm H}\beta<2.85$ 
for integrated spectra and ${\rm H}\alpha/{\rm H}\beta>2.85$ 
for nuclear spectra.  A Balmer decrement $<2.85$ 
results from a combination of: (1) intrinsically low 
reddening, (2) errors in the stellar absorption correction, and (3) 
errors in the line flux calibration and  measurement.  Errors
in the stellar absorption correction and flux calibration are discussed in
detail in Paper I, and are $\sim$12-17\% on average, with a maximum
error of $\sim30$\%.
For the S/N of our data, the lowest E(B-V) measurable is 0.02.  We therefore
assign these galaxies an upper limit of E(B-V)$<0.02$.  The difference
between applying a reddening correction with an E(B-V) of 0.02 and 0.00 
is minimal: an E(B-V) of 0.02 corresponds to an attenuation factor of 1.04
at \Ha\ using the CCM curve.

\section{Flux Coverage \label{flux}}

\subsection{Reference flux: $B_{26}$ isophote}

\begin{figure}
\plotone{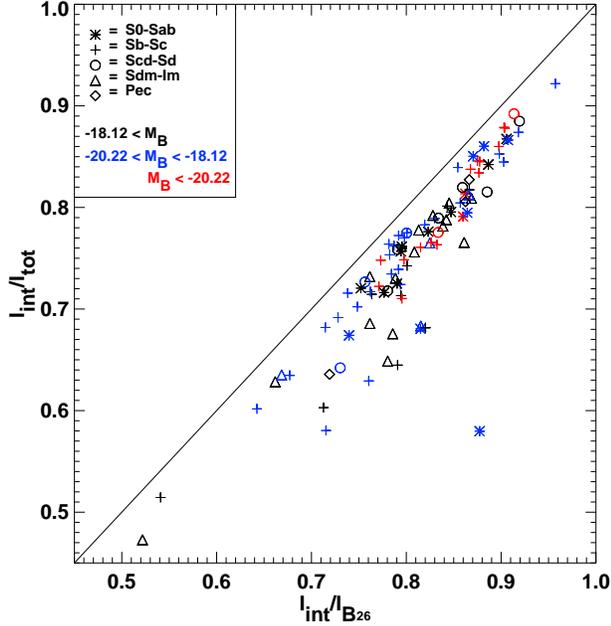}
\caption{Integrated light as a fraction of the $B_{26}$ isophotal light 
($I_{int}/I_{B_{26}}$)  
versus the integrated light as a fraction of the total B-band light 
($I_{int}/I_{tot}$).  Symbols and colors give the Hubble type and 
luminosity range shown in the legend.  We assume M$_{*}=-20.22$ 
\citep[][after conversion to our adopted cosmology]{Marzke98} .
The solid y=x line shows where 
the data would lie if $I_{B_{26}}$ is representative of the total B-band flux.  The mean offset from the y=x line is $0.039\pm0.003$.  The majority ($72/82 \sim 88$\%) of the galaxies have a mean offset close
to 0.039, regardless of luminosity.  The light contained within the B26 isophote provides a well-defined and largely unbiased proxy for the total flux.
\label{Iint_I26_Itot}}
\end{figure}

For our analysis, we define flux coverage in terms of the flux contained 
within the directly observed $B_{26}$ isophote.  Because we are investigating the 
emission-line properties of the NFGS galaxies, we assume that the majority
of the emission-line gas is contained within the $B_{26}$ isophote.
This assumption could potentially introduce
selection effects into our analysis if the flux contained within the 
$B_{26}$ isophote differs systematically from the total B-band flux as a function of 
galaxy type or luminosity. Figure \ref{Iint_I26_Itot} compares the fraction 
of $B_{26}$ light captured within our integrated spectral aperture, 
${I_{int}/I_{B_{26}}}$, with the
fraction of extrapolated total B-band light, $I_{tot}$, captured within our integrated spectral aperture, ${I_{int}/I_{tot}}$.  The solid y=x line shows where the data 
would lie if the $B_{26}$ flux represented 100\% of the B-band total flux.  
The mean deviation of our data from the y=x line is $0.039\pm0.003$.  A small fraction of galaxies (10/101) deviate from the y=x line by
more than $0.07$.   These 10 galaxies include
2 early-type spirals (S0-Sab), 5 late-type spirals 
(Sb--Sd) and 3 very late-types (Sdm--Im--Pec). 
Therefore, the use of the 
$B_{26}$ flux as a reference should not introduce a bias as a function 
of galaxy type.   The galaxies in Figure \ref{Iint_I26_Itot} are color-coded according to 
their B-band Magnitude.  We assume M$_{*}=-20.22$ 
\citep[][; after conversion to our adopted cosmology]{Marzke98}.
The 10 outlying galaxies fall in the lower two luminosity
bins ($M_{B}>-20.22$; black and blue symbols).  
However the majority ($72/82 \sim 88$\%) of the 
galaxies  with $M_{B}>-20.22$ display a similar deviation from the y=x line 
than the $M_{B}<-20.22$ galaxies (red symbols).  We conclude that the 
use of the $B_{26}$ flux as a reference should not introduce a significant
bias as a function of luminosity.  

\epsscale{1.2}

\noindent

\begin{figure}
\plotone{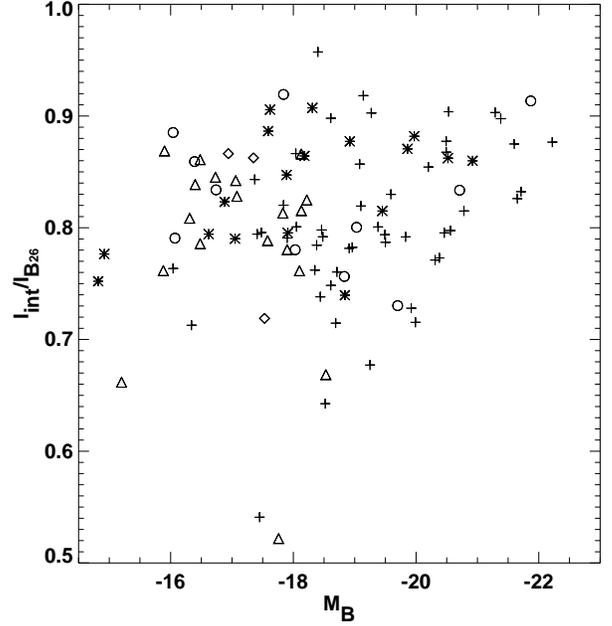}
\caption{The fraction 
of $B_{26}$ light captured within our integrated spectral aperture compared
with the blue absolute magnitude, $M_B$.  Symbols are the same as in 
Figure~\ref{Iint_I26_Itot}.  Excluding the 6 galaxies with \Iint/\I26$<0.7$, we find no
significant dependence of the integrated \B26\ fraction on absolute magnitude.
\label{Iint_I26_MB}}
\end{figure}

In Figure \ref{Iint_I26_MB}, we compare the fraction 
of $B_{26}$ light captured within our integrated spectral aperture 
(${I_{int}/I_{B_{26}}}$) with 
the blue absolute magnitude.  The Spearman Rank test gives a 
correlation coefficient of -0.19.   The probability of obtaining this value by 
chance is 6\%, indicating that
${I_{int}/I_{B_{26}}}$ is correlated (albeit weakly) with $M_{B}$.   Robust correlation methods
give similar correlation coefficients; both the Biweight Midcovariance method and 
Pearson's Product Moment Correlation give correlation coefficients of -0.22.  This correlation is driven by
the 6/101 galaxies with ${I_{int}/I_{B_{26}}} < 0.7$ and the 7/101 galaxies more 
luminous than $M_{B} = -21$.  Excluding these 13 galaxies gives a correlation
coefficient of -0.05, with a probability of obtaining this value by chance of 66\%.
The integrated spectral properties of the 6 galaxies with ${I_{int}/I_{B_{26}}} < 0.7$ 
may not be representative of their global properties.  We therefore mark these galaxies 
with a large circle in the Figures in the subsequent sections.

\begin{figure}
\plotone{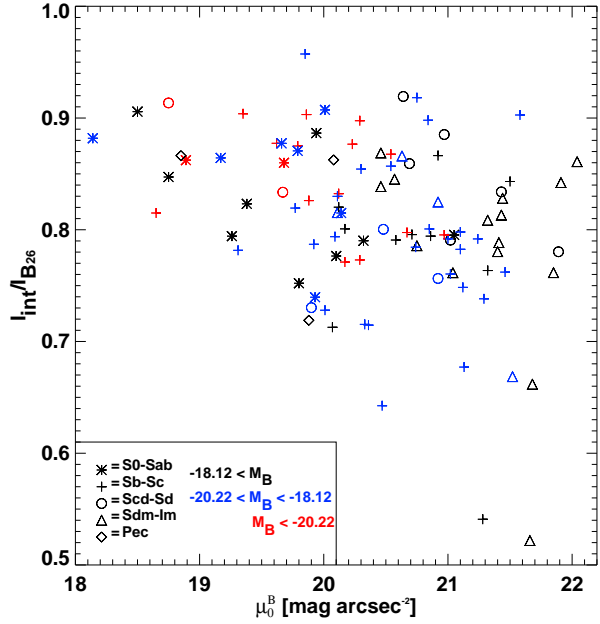}
\caption{The fraction 
of $B_{26}$ light captured within our integrated spectral aperture compared
with the B-band central surface brightness $\mu^{\rm B}_{0}$ [mag arcsec$^{-2}$]. 
We find a mild trend towards lower average integrated \B26\ fractions and a larger range in 
integrated \B26\ fractions for lower surface brightness galaxies.
\label{Iint_I26_u0}}
\end{figure}

In Figure~\ref{Iint_I26_u0} we compare ${I_{int}/I_{B_{26}}}$ with 
the B-band central surface brightness, $\mu^{B}_{0}$.  Not surprisingly, 
the integrated spectral aperture includes a larger portion of the 
$B_{26}$ light for brighter galaxies.  The Spearman Rank, Pearson's Product Moment, 
and the Biweight Midsector tests indicate 
that ${I_{int}/I_{B_{26}}}$ is correlated with $\mu^{B}_{0}$; the correlation coefficient is 
-0.36, -0.36, -0.34 respectively (the probability of obtaining this value by chance is 0.03\%).
This correlation is partly driven by the 6 galaxies with ${I_{int}/I_{B_{26}}} < 0.7$ and 
the 12 galaxies with $\mu^{B}_{0} < -19.5$.  
Removing these 17 galaxies does not entirely remove this correlation; a weak correlation remains with 
a correlation coefficient of -0.17, -0.13, -0.13 for the three methods respectively (the probability of obtaining such a value by chance is 11\%).     
Nonetheless, Figure~\ref{Iint_I26_u0}  shows that this correlation is less a
systematic  trend with $\mu_0^B$ than an increased scatter in \Iint/\I26\ toward
lower  surface brightness galaxies, such that the integrated aperture is
more likely to be representative of the total $B_{26}$ light for galaxies with the highest central 
surface brightness.

\begin{figure}
\plotone{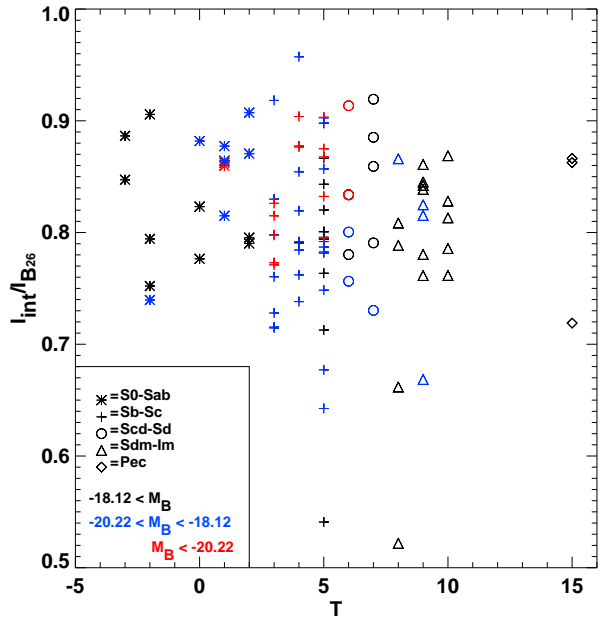}
\caption{The fraction 
of $B_{26}$ light captured within our integrated spectral aperture compared
with numeric morphological type $T$ from \citep{Jansen00a}.  
 We find no  statistically significant dependence of \Iint/\I26\ on morphological
type. \label{Iint_I26_type}}
\end{figure}

Figure \ref{Iint_I26_type} shows the range in  ${I_{int}/I_{B_{26}}}$ as 
a function of numeric morphological type, $T$, described in \citep{Jansen00a}.  There is no statistically significant systematic offset between the ${I_{int}/I_{B_{26}}}$ values as a function of numeric type.

\subsection{Nuclear versus Integrated Flux Coverage}

\begin{figure}
\plotone{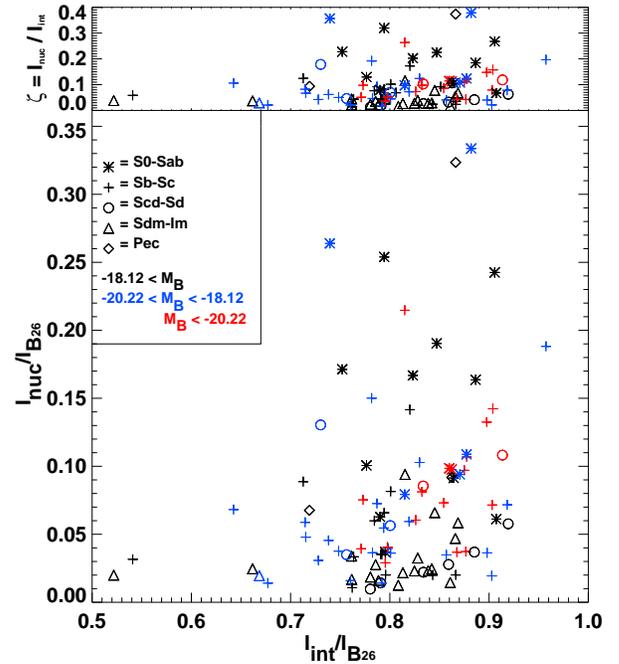}
\caption{Bottom Panel: Comparison between the nuclear and integrated 
$B_{26}$ flux covering fractions 
for the 101 NFGS galaxies in our sample.  Symbol shapes and colors are 
encoded according to 
Hubble type and 
luminosity range shown in the legend.  The average integrated 
flux covering fraction is $\sim0.81$ nuclear slit flux covering fraction 
is $\sim0.07$.  The rms scatter for both is $0.07$. 
 Top Panel:  The relative nuclear/integrated $B_{26}$ covering fraction 
($\zeta$) versus the integrated $B_{26}$ covering fraction.  The
relative $B_{26}$ covering fraction ranges between 0-0.4.  
\label{cov_frac}}
\end{figure}

Figure~\ref{cov_frac} compares the range in the fraction of \B26\ light
captured by the nuclear aperture, \Inuc/\I26, to the fraction of \B26\
light contained within the integrated aperture, \Iint/\I26, for our
101-galaxy sample.  The nuclear spectra capture a significantly
larger fraction of the \B26\ light for early- (S0--Sab) than for
late-type (Sb--Scd) spiral galaxies.  The nuclear \B26\ fraction,
\Inuc/\I26, ranges between 0.01 and 0.33, with an average value of
\tsim0.07 and an rms scatter of 0.07\/.  The integrated spectra sample
a much larger fraction of the \B26\ light: \Iint/\I26 spans the range
0.52--0.96 with an average of 0.81 and an rms scatter of 0.07\/.
For our sample, the average size\footnote{Throughout, we will express
the size of an ellipse with semi-major and -minor axes $a$ and $b$ in
terms of the \emph{elliptical} or \emph{equivalent} radius, defined as
$r=\sqrt{ab}$.} of the elliptical isophote that encompasses 81\% of the
\B26\ emission is $21\farcs94\pm0\farcs40$ (with an rms of $9\farcs60$),
or \tsim4.35~kpc\footnote{Note that the physical galaxy sizes have been 
averaged rather than a mean angular size combined with a mean redshift.}.

For the purpose of our analysis, we define the relative nuclear/integrated
$B_{26}$ fraction as:

\begin{equation}
\zeta = \frac{(I_{nuc}/I_{{B}_{26}})}{(I_{int}/I_{B_{26}})}=
\frac{I_{nuc}}{I_{int}}
\end{equation}

\noindent
The integrated spectra sample 81.0\% of the $B_{26}$
emission on average.  Hence the relative nuclear/integrated \B26\ fractions $\zeta=0.1$, 0.2, 0.3
or 1.0 correspond to 10\%, 20\%, 30\% or 100\% of \tsim81\% of the
light encompassed within the \B26\ isophote on average, or equivalently,
\tsim8.1\%, 16.2\%, 24.3\%, or 81.0\% of the total light within the \B26\
isophote.  The mean galaxy diameters for our sample that correspond to these
$\zeta$ are \tsim1.4, 2.1, 2.8 and 8.7~kpc.

\noindent
Table \ref{Type_properties}  gives the mean $\zeta$ and the mean diameter of the
\B26\ isophote for early-type spiral galaxies, late-type spirals, and
late-type/irregular galaxies (Sdm--Im/Pec). The mean value of $\zeta$
for early-type spirals is 0.17, compared with 0.07 for the later types.
In our sample, early-type galaxies are on average smaller in angular
extent than late-types and their surface brightness profiles are steeper
\citep[see][]{Jansen00a}.  Later-type galaxies have shallower profiles,
consistent with smaller contributions from a bulge component relative to
their disk \citep[e.g.,][]{Mollenhoff04}.  In Table~\ref{radius} we list
the mean elliptical radii (in arcseconds) that enclose varying fractions
of the \B26\ light, expressed both directly and in terms of the relative
nuclear/intergrated \B26\ fraction, $\zeta$.  We compute the radii from
the radial surface brightness profiles of \citet{Jansen00a}.  Because
the radii are not directly calculated from the $B$-band images, they do
not take any high spatial frequency substructure into account.

\begin{figure}
\plotone{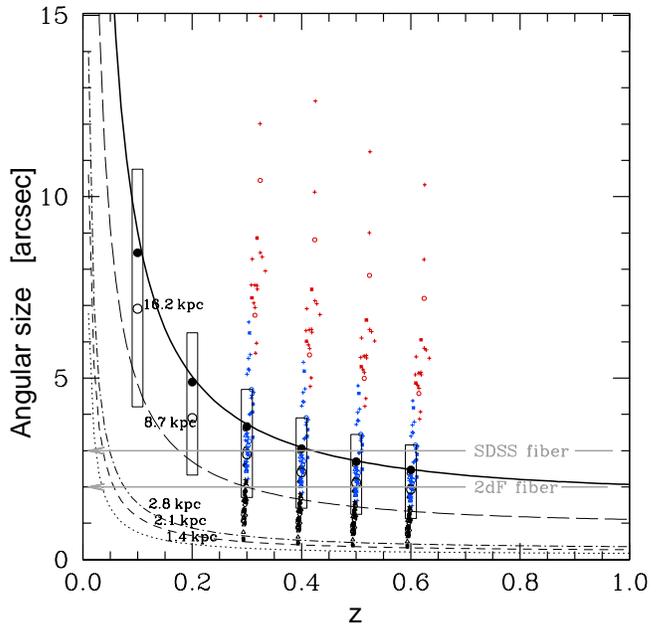}
\caption{Redshift versus angular size (arcseconds) for several values of $\zeta$.  We
artifically redshifted our 101-galaxy sample by requiring that the
median redshift become $z=0.1$, 0.2, 0.3, 0.4, 0.5 and 0.6.  The depth
of the sample along the line of sight is preserved.  The original
selection of the NFGS galaxies is reflected by the slight slant of the
data points with respect to the vertical (show only for $z\geq0.3$),
with higher luminosity galaxies (red symbols) being farther away.  
Luminosity ranges (and corresponding colors) are as in Figure~1.
Solid (open) circles represent the mean (median) angular sizes of the \B26\
isophote in the redshifted sample, and the open bars represent the
quartile range.  The solid curve, represents the mean diameter of the
\B26\ isophote (\tsim16.2~kpc for the un-redshifted sample).  Dotted,
short-dashed, dot-dashed and long-dashed curves correspond to relative
covering fractions $\zeta=0.1$, 0.2, 0.3 and 1.0, respectively.  For a
mean NFGS galaxy diameter of \tsim16.2~kpc, these values of $\zeta$
correspond to 1.4, 2.1, 2.8 and 8.7 kpc, respectively.  $\zeta=1.0$
corresponds to the average fraction of the light that is captured by our
integrated spectra, i.e., \tsim81\% of the total light enclosed by the
\B26\ isophote.  The angular sizes of the 2dF and SDSS fibers are
indicated by the grey arrows for reference.  
Reliable global metallicities require a minimum $\zeta$ of atleast 0.2 (short-dashed line).  
\label{Angsize_vs_dist}}
\end{figure}

 Figure~\ref{Angsize_vs_dist} shows the relationship between angular size
and redshift for $\zeta = 0.1, 0.2, 0.3,$ and 1.0.  
The solid line represents the mean $B_{26}$ diameter of the 
NFGS galaxies in our sample ($\sim16.2$ kpc).    
Dotted, dashed, dot-dashed,and long-dashed lines correspond to 
$\zeta = 0.1, 0.2, 0.3,$ and 1.0 respectively.  
The colored symbols on Figure~\ref{Angsize_vs_dist} show our
101-galaxy sample, artificially redshifted by requiring that 
the median redshift of the sample is at $z=0.3,0.4,0.5,0.6$.  The symbols 
give the Hubble type distribution. 
Arrows in Figure \ref{Angsize_vs_dist} mark the angular size of the 2dF and SDSS fibers.  Clearly, the $\zeta$ which would produce a spectrum that 
mimics our integrated spectrum ($\sim 81$\%) is attained with SDSS fibers 
at redshifts $z\gtrsim0.2$ and with 2dF fibers at redshifts $z\gtrsim0.3$.
 Samples containing many early-types 
require a lower redshift to obtain a high $\zeta$ with SDSS or 2dF fibers; 
samples containing many late-types require a larger redshift.  

\begin{figure}
\plotone{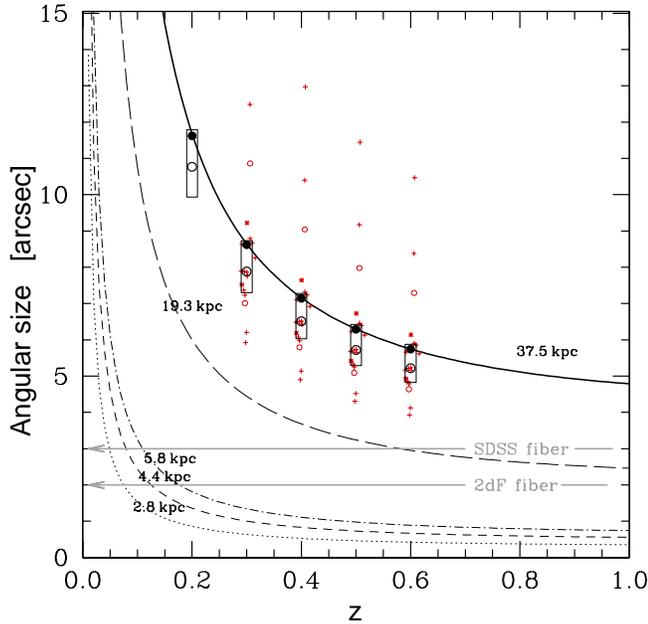}
\caption{Redshift versus angular size (arcseconds) for the high luminosity 
(${\rm M_{B}} < -20.22$) galaxies in our sample shown in Figure~\ref{Angsize_vs_dist}
The solid line represents the mean diameter ($\sim 16.2$ kpc) of the 
B$_{26}$ isophote for these high luminosity galaxies.  Dotted, short-dashed, and
dot-dashed, and long dashed lines correspond to $B_{26}$ relative covering fractions 
$\zeta = 0.1, 0.2, 0.3$ and 1.0, respectively.  These values of $\zeta$ correspond 
to diameters of 2.8, 4.4, 5.8, and 19.3 kpc.  $\zeta=1.0$
corresponds to the average fraction of the light that is captured by our
integrated spectra, i.e., \tsim81\% of the total light enclosed by the
\B26\ isophote.
The angular size of the 2dF and SDSS fibers are marked (grey arrows).
Reliable global metallicities require a minimum $\zeta$ of atleast 0.2 (short-dashed line).  
\label{Angsize_vs_dist2}}
\end{figure}

The symbol colors in Figure \ref{Angsize_vs_dist} correspond to the three
luminosity bins shown in previous figures.  There is  a strong 
dependence of $\zeta$ on luminosity.  At higher redshift, magnitude limited 
samples contain galaxies with higher luminosities on average than local 
samples because the galaxies at the faint end of 
the luminosity function are missing.  
Figure \ref{Angsize_vs_dist2} shows $\zeta = 0.1, 0.2, 0.3,1.0$ for the mean diameter of the 
$M_{B}<M_{\star}$ galaxies in 
our sample.  
Clearly for samples containing $M<M_{\star}$ galaxies similar
to those in the NFGS, there is no 
practical redshift which provides a $\zeta$ that mimics our integrated
spectrum.  Figure \ref{Angsize_vs_dist2} illustrates the importance of
understanding the relationship between the luminosity range of a sample 
and the flux coverage of the aperture.  In the following Sections,
we investigate the flux coverage required to provide
representative estimates of metallicity, star formation rate and extinction.

 In the remainder of this paper, 
we assume that our integrated spectra are representative of the 
emission-line properties within the $B_{26}$ isophote.  Six galaxies 
have integrated covering fractions $<0.7$, for which this assumption 
may not be valid.  We mark these galaxies in each figure 
with a circle.

\section{Metallicity \label{Metallicity}}

Aperture effects on metallicity estimates result from metallicity
gradients.  \citet{Aller42} first noticed gradients in the 
emission-line ratio \OIIIHb\ within the 
late-type spiral M33.  \citet{Searle71} then 
interpreted the radial change in the \OIIIHb\ ratio as a radial decrease 
in metallicity (O/H).  Abundance gradients have now been observed in many 
galaxies \citep[see][for a review]{Henry99}.  Various hypotheses have 
been proposed to explain abundance
gradients: (1) radial variation of stellar yields caused by IMF 
differences between the spiral arms and the interarm regions \citep[e.g.][]{Guesten82}; (2) a star formation rate dependence on radius \citep{Phillipps91}; (3) radial infall of primordial gas during disk formation 
\citep{Matteucci89,Pagel89,Edmunds95}.  Current chemical evolution models
include a combination of these three processes 
\citep[e.g.,][]{Churches01}.

The dependence of the metallicity gradient on galaxy type is unknown.
Massive late-type spirals like M101 have large 
metallicity gradients; the \HII\ region metallicities decrease by an order 
of magnitude from the inner to the outer disk 
\citep[see e.g.,][for a review]{Shields90}.  
However, not all galaxies have strong metallicity gradients.
Many barred spiral galaxies have weaker metallicity gradients than 
spirals of similar type \citep[e.g.,][]{Pagel79,Roy97}, suggesting that 
radial gas flows suppress or mix metallicity gradients 
\citep[e.g.,]{Roberts79,Martin94,Roy97} \citep[but c.f.][]{Considere00}

Despite numerous studies, it is unclear whether early-type spiral 
galaxies have weaker metallicity 
gradients than late types.  Metallicity gradient comparisons may be 
affected by variations in mass, luminosity and gas fraction, as well as 
dynamical effects such as radial gas flows 
\citep[see e.g.,][for a review]{Molla96}. 
\citet{Smith75} found weaker gradients in 
early-type spirals than in late types.  However, \citet{Edmunds84} argued
that the metallicity gradient in early-types is similar to that in late
types.  \citet{Garnett87} showed that the Sab galaxy M81 has a similar gradient to the gradients observed in later-type spirals with similar mass, 
luminosity, and gas fraction.   

These complex metallicity gradients may introduce random and/or 
systematic errors into the luminosity-metallicity relation found 
in fiber surveys 
\citep{Tremonti04,Lamareille04,Maier04}. The combination of strong 
metallicity gradients with a small fixed-angular size 
aperture may produce a decrease in global galaxy metallicities with 
increasing redshift until the aperture size is comparable to the galaxy 
size.  On the other hand, weak metallicity gradients may have little or 
no effect on metallicities obtained with a small fixed-angular size aperture.  

\begin{figure}
\plotone{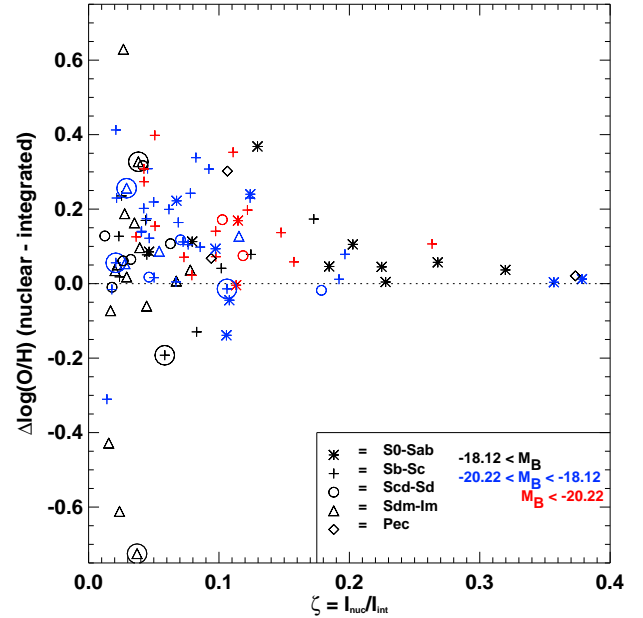}
\caption{Abundance ratio (nuclear/integrated) in log units versus the relative 
(nuclear/integrated) $B_{26}$ covering fraction $\zeta$ 
for the NFGS galaxies as a function
of Hubble type and luminosity at a fixed $\zeta$. The six galaxies 
with integrated $B_{26}$ fractions $I_{int}/I_{B_{26}}<0.7$ are marked with a large circle. 
For small $\zeta$ (0-0.1) the abundance ratio 
has considerable scatter.  The scatter is a function of galaxy type. 
With small apertures ($\lesssim 10$\% of the light) the metallicity 
is systematically over-estimated for most galaxies.  Although errors in the absolute 
metallicities are $\sim0.1$ in log(O/H)+12 units, these errors primarily 
reflect the accuracy of the theoretical models used to create the metallicity diagnostics.
Any error introduced by the diagnostics is likely to be systematic 
\citep[see][for a discussion]{Kewley02a,Kobulnicky04}.  Because we are using {\it the same} metallicity diagnostic for both nuclear 
and integrated spectra, errors in the {\it relative} metallicities should 
be $\ll 0.1$ dex.
\label{Abund_vs_cov_type}}
\end{figure}

Figure \ref{Abund_vs_cov_type} shows the
metallicity ratio (nuclear metallicity - integrated metallicity 
in log(O/H) units) versus $\zeta$ for the Hubble types and
luminosity ranges in our sample.   Use of $\zeta$ as a covering fraction 
diagnostic assumes that our integrated spectra are representative
of the total emission within the $B_{26}$ isophotal radius.  
The dotted line shows where the
data would lie if the nuclear and integrated metallicities were identical.
The nuclear metallicities exceed the integrated metallicities
 on average by $\sim0.13$ dex for our 101-galaxy sample.

\begin{figure}
\plotone{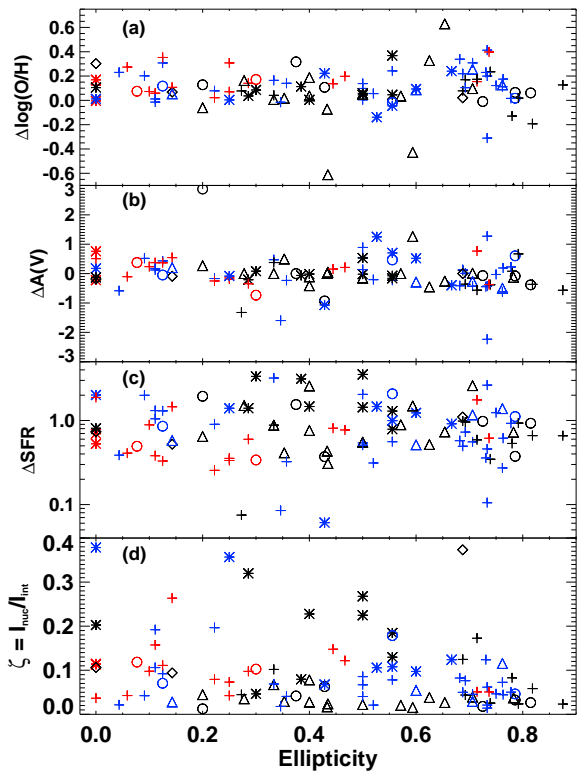}
\caption{Ellipticity from \citet{Jansen00a} versus (a) Abundance ratio (nuclear/integrated) in 
log units, (b) Nuclear - integrated extinction A(V), (c) SFR ratio (nuclear/integrated), and (d) 
relative (nuclear/integrated) $B_{26}$ covering fraction $\zeta$.  There is no clear correlation between 
ellipticity and abundance ratio, extinction ratio, SFR ratio, or covering fraction.
\label{ellipticity}}
\end{figure}

The inclination of the galaxies (whether a galaxy is viewed face-on or 
edge-on) may influence whether a strong abundance gradient is
observed.  In Figure~\ref{ellipticity}a we plot ellipticity versus metallicity ratio.  
There is no correlation between ellipticity and nuclear to integrated metallicity ratio.
Because ellipticity is only a very crude measure of inclination, a lack of 
correlation between ellipticity and metallicity ratio does not rule out a correlation 
between inclination and metallicity ratio.

For small $\zeta$
(0-0.1), the metallicity ratio has a large scatter.
Table \ref{Type_properties} gives the mean and rms scatter in $\zeta$
and metallicity ratio for early type spirals (S0-Sab), 
late type spirals (Sb-Sd) and very late Hubble types (Sdm-Im/Pec).  
Clearly the galaxy distribution in Figure \ref{Abund_vs_cov_type} is
influenced by Hubble type and luminosity. 
The mean metallicity ratio for late types (Sb-Sd) 
is $0.14\pm0.02$ dex compared with $0.09\pm0.03$ dex for the early 
types (S0-Sab) and $0.11\pm0.06$ dex for the very late types (Sdm-Im-Pec).  
We note that the NFGS early-type galaxies 
have larger $\zeta$ on average than the later galaxy 
types.  The minimum $\zeta$ for early-types is 0.046.  
For $0.046<\zeta<0.15$, the difference between 
the early, late-type and very late-type mean metallicity ratios is not 
significant within the errors ($0.13\pm0.05$, $0.14\pm0.02$, $0.11\pm0.05$).

Figure \ref{Abund_vs_cov_type} and Table \ref{Type_properties} show that 
the very late-type galaxies show 
considerably more scatter in metallicity ratio 
than early types or late-types.  This difference is dominated by the 4 Sdm-Im galaxies with 
small $\zeta$; for $0.046<\zeta<0.15$, 
the early-types and late types have comparable scatter about their mean metallicity 
to the very late-types (0.13-0.14 c.f. 0.10).   However, a larger sample of early-types is required 
to judge whether the difference in scatter at small $\zeta$ is significant.   
Table \ref{Type_lum} gives the fraction of galaxies of each Hubble type within three luminosity ranges.  
The Sdm, Im, and Peculiar galaxies have the smallest range in luminosities,
with the majority of luminosities ${\rm M_B}>-18.12$.  Low-luminosity 
Sdm, Im, and Peculiar galaxies often have diffuse or patchy star-formation 
distributions that may result in a wide range of metallicity gradient 
slopes \citep[see Figure 2.3 in][]{Jansen00a}.   A patchy metallicity distribution might result 
from randomly distributed star forming regions with a relatively long time between bursts.

\subsection{Effect of Aperture on Metallicity}

Figure \ref{Abund_vs_cov_type} and Table \ref{Type_properties} show that if 
$\zeta \lesssim 0.2$, the estimated metallicity exceeds the global metallicity by 
$\sim 0.13$ dex (log(O/H) units) on average.  The range in metallicity 
gradients observed 
in galaxies (i.e., the scatter in Figure \ref{Abund_vs_cov_type}) 
introduces an error up to $\pm0.6$ dex in 
metallicity.  As $\zeta$ increases from 0.01
to 0.4, the scatter decreases and the metallicity ratio tends towards
zero.  The data suggest that a flux covering fraction of $\zeta >0.2$ (i.e. $\gtrsim 20$\% of 
the galaxy $B_{26}$ light) is required for 
a reliable indication of the global metallicity value.   
 We note that only a small fraction of our sample (9 galaxies) have $\zeta >0.2$.  
The majority (6/8) of these galaxies are early-type spirals.  A similar number of early-type
spirals in our sample have  $\zeta <0.2$ (11 galaxies).   If the reduction in scatter for $\zeta >0.2$ is not real, 
then we would expect the scatter to be similar for the early-types with $\zeta <0.2$ and the 
early-types with $\zeta >0.2$.  The rms scatter for early-types for $\zeta <0.2$ is 0.14 compared to
0.03 for $\zeta >0.2$, indicating that the reduction in scatter for $\zeta >0.2$ is likely to be real. 
For other galaxy types, our results indicate that an aperture which contains less than 20\% of the global galaxy $B_{26}$ light is inadequate.   A larger sample is required to verify the specific aperture size at which the metallicities will approximate the global values.
 
For the SDSS fiber, a redshift $z\gtrsim0.04$ is required to reach 
$\zeta \gtrsim 0.2$, assuming the 
SDSS samples cover a similar range of luminosity and Hubble type as our 
sample. For the
2dF fiber, a redshift $z\gtrsim0.06$ is required.
If metallicities are  estimated using strong-line ratios from the 
SDSS or 2dF fiber spectra 
for objects at redshifts $z\lesssim0.04$ or $z\lesssim0.06$ respectively, 
errors in the metallicity estimates could be significant (up to 
$\pm 0.6$ dex in log(O/H)), and metallicities are likely to be systematically
over-estimated by $\sim0.13$ dex on average if the late-type fraction is
similar to the NFGS.  The situation is worse for high luminosity samples.
For SDSS samples with a mean luminosity $L>L_{\star}$, a redshift of 
$z\gtrsim0.09$ is required to reach $\zeta \gtrsim 0.2$. For 2dFGRS samples, 
with $L>L_{\star}$, a a redshift of $z\gtrsim0.14$ is required. 
Note that more conservative (i.e. larger) lower limits on $z$ are 
required to ensure that most galaxies (not just the typical galaxy) in a 
given sample yield reliable metallicity estimates.

\subsection{Implications for the Luminosity-Metallicity Relation \label{ap_metallicity}}

Recent investigations into the luminosity-metallicity relation using fiber
surveys may be affected by the metallicity gradient aperture effect.
For galaxies at the same $z$, the aperture effect is larger 
at higher luminosities.  This problem may affect the slope of 
luminosity-metallicity relations.

\citet{Lamareille04} calculated the luminosity-metallicity relation 
for 6387 2dFGRS galaxies between redshifts of $0<z<0.15$, with a median
redshift of 0.11.  The redshift limit of
$z=0.15$ corresponds to our $B_{26}$ covering fraction of $\sim50$\%
and  $\zeta \sim0.6$, assuming that 
the median size, luminosity and type of the 2dFGRS galaxies is the same as 
in the NFGS sample.  The median redshift of the 2dFGRS sample corresponds to a
$\zeta \sim0.45$.  It is likely therefore that 
a minor fraction of the 2dFGRS metallicities 
(galaxies at redshifts $\lesssim0.06$) are affected by the 
aperture-metallicity effect (Figure \ref{Abund_vs_cov_type}).  
The galaxies at redshifts $<0.06$ may have apparent metallicities 
higher than their global metallicities by $\sim0.1$ dex on average with
a large scatter introduced by aperture effects and metallicity
gradients.  

\citet{Schulte04} calculated the SDSS luminosity-metallicity relation 
 for 13,000 SDSS galaxies between redshifts of $0<z<0.3$.
They show luminosity-metallicity plots for
large, small and compact star-forming galaxies within 
four redshift ranges: $0<z\le 0.05$, $0.05<z\le0.1$, $0.1<z\le0.2$ and 
$0.2<z\le0.3$.  The 3$^{\prime \prime}$ SDSS fiber produces a
$\zeta \lesssim0.15$ for the redshift range $0<z\le 0.05$ for the typical 
diameter of our NFGS sample.  In this redshift range, the SDSS late-type 
star-forming galaxies may have metallicities that are overestimated 
compared to their global metallicities.  In addition, the metallicity 
estimates of the late-types and very-late types in the SDSS sample 
are probably prone to the large scatter that we observe in 
Figure \ref{Abund_vs_cov_type} at small $\zeta$.  The $0<z\le 0.05$ 
luminosity-metallicity relation is probably 
dominated by aperture effects.  Figure \ref{Angsize_vs_dist} indicates
that the remaining redshift ranges in Schulte-Ladbeck et al. are probably
not strongly influenced by aperture effects except possibly at the highest 
luminosities, unless their galaxy properties differ significantly from those
in the NFGS, and/or our chosen minimum covering fraction of $\zeta > 0.2$ is too low for 
late type galaxies that are not well represented in our $\zeta > 0.2$  sample.

 \citet{Tremonti04} used 210,000 SDSS galaxies with redshifts between 
$0.03<z<0.25$ to calculate the luminosity-metallicity and mass-metallicity relations.  
They selected this redshift range to minimize aperture effects.  
As a result of
this selection criterion, the SDSS luminosity-metallicity and mass-metallicity relations are
probably not seriously affected by a metallicity gradient aperture effect.

We note that luminosity-metallicity relations are very sensitive to the 
choice of metallicity diagnostic; different diagnostics can produce 
metallicity estimates that differ systematically by 
$\sim 0.1 - 0.2$ dex \citep[e.g.,][]{Kewley02b}.  Therefore, differences 
between \citet{Lamareille04} and other luminosity-metallicity relations 
may be influenced by the choice of metallicity diagnostic as well as 
by the aperture-metallicity effect.

We conclude that comparison of luminosity-metallicity relations requires 
(1) careful attention to aperture effects, particularly to the dependence of these effects on luminosity, and (2) use of the same metallicity diagnostic.

\section{Extinction \label{Extinction}}

Dust grains form from heavy elements.  Thus one might expect 
an extinction gradient in disk galaxies  similar to 
the metallicity gradient (unless the dust grains are depleted from 
the \HII\ regions).  Observations of extinction in \HII\ regions of
disk galaxies give conflicting results.
\citet{Sarazin76} used radio and optical 
observations to show that extinction gradients exist in M101 and 
M33.  These gradients were confirmed by \citet{Viallefond86} and
\citet{Israel80}.  More recently,  \citet{Jansen94} and \citet{Roy97} 
reached similar conclusions.  \citet{Jansen94} showed that 
similar extinction gradients exist in three highly-inclined spiral galaxies
to those in our Galaxy and in the Sombrero galaxy.
\citet{Roy97} examined the extinction gradient in the barred spiral galaxy 
NGC 1365 and showed that the extinction decreases 0.4-0.6 mag from the 
nucleus to the edge of the disk.
However, \citet{Kaufman87} found a significant scatter in radio-derived 
extinction measurements for \HII\ regions in the disk of the 
early-type spiral galaxy M81.  Kaufman et al. ascribed this scatter 
to a patchy dust distribution external to the \HII\ regions.
They found little evidence for a strong extinction gradient, 
in contrast with the previous work on M33 and M101.  We note that M81 is a 
member of an interacting triplet and its extinction properties may be 
influenced by the interaction.
However, \citet{Kennicutt89} reached conclusions similar to Kaufman et al. 
in their comparison between various \HII\ region-like nuclei and disk 
\HII\ regions.  Kennicutt et al. tentatively conclude that although 
optical extinction 
(derived from the Balmer decrement) is important for individual \HII\ 
regions and nuclei, the effect of extinction between nuclear and disk 
\HII\ regions is similar.

\begin{figure}
\plotone{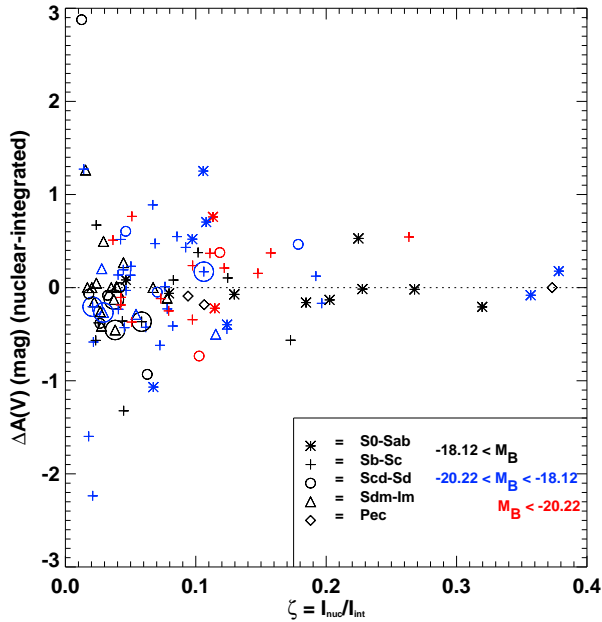}
\caption{Nuclear - integrated extinction A(V) versus the relative 
(nuclear/integrated) B$_{26}$ covering fraction $\zeta$ 
for the NFGS galaxies as a function
of Hubble type.  The six galaxies 
with integrated $B_{26}$ fractions $I_{int}/I_{B_{26}}<0.7$ are marked with a large circle. 
For small $\zeta$ (0.01-0.1) $\Delta$A(V)
has considerable scatter.  The estimated error in $\Delta$A(V) is $\pm0.09$.
\label{AV_vs_cov_type}}
\end{figure}

Figure \ref{AV_vs_cov_type} shows the nuclear - integrated $A(V)$ versus 
$\zeta$.  Table \ref{Type_properties} gives
the mean $\Delta A(V)$ and the scatter about the mean for the
early, late, and very late type galaxies.  As we have discussed in 
Section 2, a number of 
galaxies in our sample have intrinsically low reddening, with an upper 
limit of $E(B-V)<0.02$ ($A(V)=0.061$) for their integrated and/or 
nuclear spectra.  We
calculated the mean and rms $\Delta A(V)$  for $A(V)=0.061$ and 
$A(V)=0.0$.  
The difference between using  $A(V)=0.061$ or 
$A(V)=0.0$ is smaller than the size of the data points 
in Figure \ref{AV_vs_cov_type} and the effect on  $A(V)$ properties in 
Table \ref{Type_properties} is negligible.

In Figure~\ref{ellipticity}b we plot ellipticity versus $\Delta A(V)$.  
There is no correlation between ellipticity and $\Delta A(V)$ but this result
 does not rule out a correlation between inclination and $\Delta A(V)$.

The sample mean $\Delta A(V)$ is consistent with $\sim0$ within the errors, 
indicating that there is no significant offset between the nuclear and integrated $A(V)$ 
on average.  
This result is consistent with the flat extinction gradients 
observed by \citet{Kennicutt89} and \citet{Kaufman87}, but our result 
appears to be inconsistent with the extinction gradients observed with
radio data for M101 and M33.  However, radio-derived extinction measurements are known to 
systematically exceed those calculated using the Balmer decrement
 for the same
\HII\ regions \citep[e.g.,][]{Sarazin76,Israel80,Caplan86,vanderHulst88,
Kennicutt89, Bell02}.  Indeed, radio continuum fluxes predict 
a higher extinction in \HII\ nuclei than in disk \HII\ regions, even though
the extinction measured using the Balmer decrement is similar
 (with large scatter) for nuclear and disk \HII\ regions 
\citep{Kennicutt89}.   Kennicutt et al. attributed this discrepancy to 
low-level AGN contribution.  However a similar discrepancy is seen in 
the Magellanic Clouds HII regions study by \citet{Bell02}.  The cause of this
discrepancy is unknown.  A lower Balmer decrement extinction estimate 
may result from the
presence of dust grains with a flat 
optical absorption curve \citep{Sarazin76}, 
or an inhomogeneous (clumpy) distribution of absorbing material/dust 
\citep{Sarazin76,Caplan86,Kaufman87,vanderHulst88}.   The presence of 
clumpy dust may cause scatter in Figure~\ref{AV_vs_cov_type}.  Systematic 
effects on the optical Balmer decrement are likely to be minimized in 
Figure~\ref{AV_vs_cov_type} because we use the ratio of the 
nuclear to integrated Balmer decrement where both the nuclear and integrated 
Balmer decrements are derived using the same method.

\subsection{Effect of Aperture on Extinction}

Figure \ref{AV_vs_cov_type} indicates that for  $\zeta <0.2$,
the extinction for an individual galaxy 
may be overestimated or underestimated by up to 1 mag.  If $\zeta$
 is very small ($\lesssim 0.03$), the extinction may be up to 
3 magnitudes smaller or larger than the global value.  Late-type galaxies
are probably more strongly affected by this aperture effect than early
or very late types; late type galaxies have the largest scatter in 
$\Delta A(V)$ (0.7 rms vs. 0.4-0.5 rms).  
The best global extinction estimates from the 2dFGRS and SDSS are likely to occur 
at redshifts $\gtrsim0.06$ and redshifts $\gtrsim0.04$, respectively
 (Figure \ref{Angsize_vs_dist}) for samples with similar properties to
the NFGS.  Samples containing higher luminosity galaxies, or many late-type 
galaxies require larger minimum redshift limits.

\section{Star Formation Rate \label{SFR}}

Metallicity and extinction estimates are usually calculated using 
emission-line ratios that have not been corrected for metallicity 
or extinction gradients because these gradients are not known a priori.  
Optical star formation-rates, on the other hand, require 
emission-line luminosities.  The effect of aperture on emission-line 
luminosities can lead to strong biases in star-formation rate studies.
The potential implications of these biases are discussed by several authors
\citep[][]{Gomez03,Perez03,Hopkins03,Nakamura03}.  

To avoid strong biases, the luminosities and star-formation rates 
derived from fiber surveys 
are corrected for aperture effects.  The 
simplest aperture correction methods assume that the emission measured 
through the fiber is representative of the entire galaxy.  One such 
method was applied to the SDSS \Ha\ and \OII\ SFRs by \citet{Hopkins03} 
(hereafter H03).  
H03 multiplied the emission-line equivalent widths by the 
flux at the effective wavelength of the SDSS $r$-band filter.   
H03 also applied a more explicit aperture
correction $A$ based on the difference between the total galaxy 
magnitude ${\rm r_{total}}$ and the magnitude calculated from the 
fiber spectrum ${\rm r_{fiber}}$:

\begin{equation}
A=10^{-0.4({\rm r_{total}}-{\rm r_{fiber}})}.
\end{equation}

This technique is roughly equivalent to multiplying our NFGS nuclear 
fluxes by the ratio of the integrated to nuclear covering fractions.  
Both of the methods outlined in H03 assume that the distribution of the line 
emission (active star formation regions) is identical to that of the 
stellar continuum emission (older stellar population). However.  \Ha\ and 
$r$-band continuum images of spiral galaxies show marked
differences.  \citet{Lehnert96} showed that 
spiral galaxies selected to have regular 
symmetric profiles in $r$-band continuum images do not have regular,
symmetric \Ha+\NII\ emission.  Rather, the \Ha+\NII\ emission appears 
clumpy with filaments.  In some cases, the \Ha+\NII\ emission forms 
large-scale regions with bubble and/or shell-like structures.  
Even early-type spirals show very diverse \Ha\ morphologies, despite 
similar $r-$band continuum morphologies \citep[][]{Hameed99}.

\citet{Brinchmann04} proposed a more complex aperture correction method.
They calculate the likelihood 
distribution of the star formation rate for a given set of spectra with 
global and fiber $g-r$ and $r-i$ colours.  
Brinchmann et al. tested this method using galaxies within the SDSS
with different fiber covering fractions.  They conclude that their 
correction method is robust only if $\ge 20$\% of the total $r$-band light
is sampled by the fiber, in remarkable agreement with our conclusions
for metallicity and extinction.

We calculate star-formation rates for our NFGS sample using the \Ha\ 
luminosities and the \citet{Kennicutt98} SFR(\Ha) calibration 
\citep[as described in][]{Kewley02a}. To simulate the effect of the simple 
aperture correction methods of H03, we multiply our nuclear SFRs by the 
ratio of the 
corresponding integrated to nuclear covering fractions.  This process 
provides crude `expected' global SFR estimates for comparison 
with SFRs derived from our integrated spectra.

In Figure~\ref{ellipticity}c we plot ellipticity versus  the 
ratio of `expected' to integrated \Ha\ star formation rates (SFR ratio).
There is no correlation between ellipticity and SFR ratio but this result 
does not rule out a correlation between inclination and SFR ratio.

\begin{figure}
\plotone{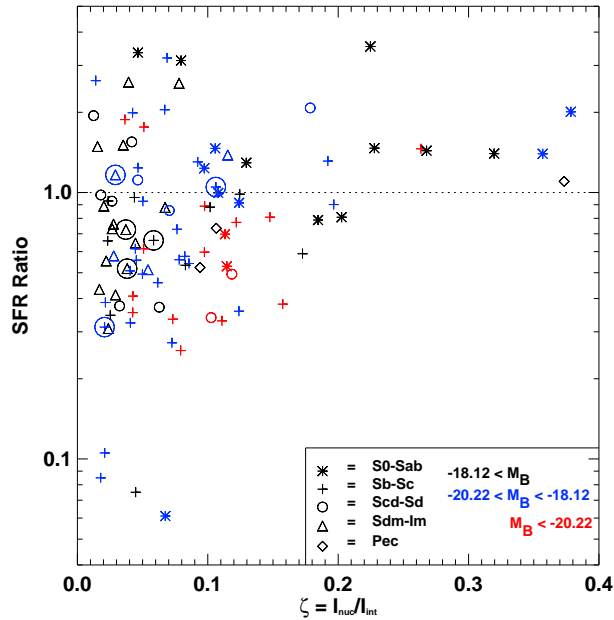}
\caption{SFR ratio (nuclear/integrated) versus the relative 
(nuclear/integrated) $B_{26}$ covering fraction $\zeta$ 
 for the NFGS galaxies as a function
of Hubble type.  The six galaxies 
with integrated $B_{26}$ fractions ${\rm I_{int}/I_{B_{26}}}<0.7$ are marked with a large circle. 
 For small $\zeta$ (0.01-0.2) the SFR ratio 
has considerable scatter.  The scatter is a function of galaxy type. The estimated error in 
the SFR from \Ha\ for the NFGS is $\sim 30\%$.
\label{SFR_int_nuc_cov_type}}
\end{figure}

Figure \ref{SFR_int_nuc_cov_type} shows the 
ratio of expected to integrated \Ha\ star formation rates versus 
$\zeta$.   Table \ref{Type_properties} gives the mean and 
rms scatter for the early, late and very late types. The mean SFR ratio for  
early-type spirals is $1.47\pm0.23$.  A mean SFR ratio $>1$ indicates that 
 SFRs calculated by this widely used prescription 
overestimate the true global SFRs on average, 
despite the fact that the early types have 
a larger $\zeta$ (0.17 c.f. 0.07).  H03 found a similar
effect when they compared their \Ha\ SFRs with those measured using the
radio 1.4GHz flux.  The H03 \Ha\ SFRs are overestimated for 
galaxies requiring the largest aperture corrections.  The largest
galaxies typically require the largest aperture corrections.  
The expected SFRs
may be overestimated if the star-forming regions in early-types are
centrally concentrated \citep{Hameed99,James04}.

The mean SFR ratio for the late-types is $0.85\pm0.08$, 
implying that  we miss $\sim10-20$\% of the star formation 
by assuming that the emission-line gas
follows an identical profile to the continuum emission.  
Star formation that is extended or that occurs preferentially in the outer 
regions of late type galaxies probably explains this result 
\citep[e.g.,][]{Bendo02}.  

For small $\zeta<0.2$, there is substantial scatter 
between the nuclear and integrated SFR measurements (rms scatter $=0.70$).  We therefore
recommend that SFR studies use spectra with $\zeta >0.2$,
confirming the conclusion of \citet{Brinchmann04}.

For $\zeta>0.2$, although the scatter is reduced, 
there is a danger that the expected SFR value over-estimates the 
true SFR.  For $\zeta >0.2$ the corrected small aperture 
measurement 
predicts a value that is $60\pm20$\% larger than our integrated spectrum, 
even when the aperture $\zeta$ approaches 0.4.  Only 8 galaxies 
in our sample have  $\zeta>0.2$ and most of these galaxies
are early-types.  A larger sample of 
galaxies with flux covering fractions between 0.2-1 is required to determine
whether this SFR overestimate is a serious concern.  However, these
results serve as a warning that there is danger in determining the 
global SFR by applying a 
simple aperture correction, even for $\zeta$ as large as 0.4.
 Note that in a fixed aperture fiber survey galaxies with higher luminosities or late-type galaxies
need to be at a larger redshift to ensure that $\zeta >0.2$.

\section{Implications for Star Formation History Studies}

Some fiber-based SFR studies cover redshifts $z>0.05$ 
to minimize the potential aperture bias \citep{Gomez03,Brinchmann04}.  
As we discussed in Section \ref{ap_metallicity}, a redshift of $z>0.05$ 
corresponds to mean $\zeta >0.2$ for the SDSS.  
Figure \ref{SFR_int_nuc_cov_type} and the results of Brinchmann et al. 
show that  $\zeta >0.2$ is sufficient to avoid the strong 
scatter  introduced with smaller flux covering fractions.  However, 
SFR calculations that assume identical \Ha-emission and continuum
distributions may systematically overestimate  SDSS SFRs 
up to redshifts of at least $z=0.1$ (assuming a 17.7 kpc $B_{26}$ isophotal 
diameter galaxy), especially if the galaxies contain star-forming 
regions that are more 
centrally concentrated than the continuum emission.   
Care should be taken when comparing star-formation rates derived from 
emission-lines in fiber surveys and those obtained with larger apertures 
or by photometric methods.

\section{Conclusions \label{Conclusions}}

We investigate the effect of aperture size on the star formation rate,
metallicity and extinction for 101 galaxies from the objectively
selected Nearby Field Galaxy Survey.  Our sample includes 
galaxies of all Hubble types except ellipticals with SFRs ranging from 0.01 to 100 
\Msun yr$^{-1}$ and spans a large range in metallicity, star-formation rate
and extinction.  We calculate an `expected' star formation rate using 
 nuclear spectra and applying the commonly-used aperture correction 
method.  We compare the metallicity, extinction and star formation rate 
derived from nuclear spectra to those 
derived from integrated spectra.   We find that:

\begin{itemize}
\item  For flux covering fractions $<20$\% ($\zeta<0.2$)
the difference between the nuclear and global metallicity, extinction and
star formation rate is substantial.  An aperture that contains less than 20\% of the  $B_{26}$ light is inadequate at providing reliable estimates of the global metallicity, SFR or extinction.    A larger sample of galaxies is required to determine the specific aperture size at which the metallicity, extinction, and star formation rates will approximate the global values.

\item For flux covering fractions $<20$\%, the aperture effect on 
metallicity depends on Hubble type.  This dependence occurs because of 
differences in the metallicity gradients.  Late-type spiral galaxies 
show the largest systematic difference ($\sim0.14$ dex) between nuclear 
and global metallicities.  The largest difference (scatter) between the nuclear 
and global metallicities occurs for Sdm--Im, and Peculiar galaxies, 
indicating a large range in metallicity gradients for these galaxy types.  

\item We find little evidence for systematic differences between nuclear 
and global extinction estimates for any galaxy type.  However, 
there is significant scatter between the nuclear and integrated 
extinction estimates for flux covering fractions $<20$\%.

\item The `expected' star formation rate overestimates the global value 
for early-type spirals and a slightly underestimates the global value 
for late-type spirals, but with large scatter.  The systematic differences
probably result from the aperture correction assumption
that the distributions of the emission-line gas and the continuum are 
identical.  The large scatter (error) in the estimated SFR occurs 
when the aperture covering fraction is $<20$\%. 
\end{itemize}

These aperture effects impact investigations into the luminosity-metallicity 
relation and star-formation history 
using fiber surveys unless samples have been 
selected with a lower limit on $z$.  To avoid the systematic and random 
errors from aperture effects, we recommend selecting samples with fiber 
covering fractions that capture $>20$\% of the galaxy light.  
Redshifts 
$z>0.04$ and $z>0.06$ are required to minimize these aperture effects in the 
Sloan Digital Sky Survey and the 2dF Galaxy Redshift Survey respectively, assuming that these
samples contain galaxies with similar properties to those in the NFGS.  
Samples containing galaxies with higher luminosities and samples containing 
more late-type galaxies than in the NFGS require larger redshift limits.

\acknowledgments
We wish to thank the anonymous referee for his/her careful reading of the manuscript and 
useful comments.  L. J. Kewley is supported by a Harvard-Smithsonian 
CfA Fellowship.   M. J. Geller is supported by the Smithsonian Institution.

\newpage

\LongTables
\begin{deluxetable}{rllllll}
\tablecolumns{7}
\tablewidth{10cm}
\tabletypesize{\scriptsize}
\tablecaption{Classification and Quality Flags for the NFGS
\label{classtable}}
\tablehead{ID & Name
& \multicolumn{2}{c}{Integrated}
& \multicolumn{2}{c}{Nuclear}
& Adopted\\
 & & Class\tablenotemark{a} & Quality\tablenotemark{b} & Class\tablenotemark{a} & $Q$\tablenotemark{b} & Class\tablenotemark{a}\\
}
\startdata
          1 & A00113+3037 & \nodata & \nodata & \nodata & \nodata & ... \\
           2 & A00289+0556 & HII & 1 & HII & 1 & HII \\
           3 & NGC 193 & \nodata & \nodata & \nodata & \nodata & ... \\
           4 & A00389-0159 & HII & 1 & HII & 3 & HII \\
           5 & A00442+3224 & HII & 1 & HII & 3 & HII \\
           6 & A00510+1225 & Sy1 & 1 & Sy1 & 1 & AGN \\
           7 & NGC 315 & \nodata & \nodata & AGN & \nodata & AGN \\
           8 & A00570+1504 & AGN & 1 & AGN & 1 & AGN \\
           9 & A01047+1625 & HII & 1 & \nodata & \nodata & HII \\
          10 & NGC 382 & \nodata & \nodata & HII & 3 & HII \\
          11 & IC 1639 & \nodata & \nodata & \nodata & \nodata & ... \\
          12 & A01123-0046 & Sy1 & 1 & \nodata & \nodata & AGN \\
          13 & A01187-0048 & \nodata & \nodata & \nodata & \nodata & ... \\
          14 & NGC 516 & \nodata & \nodata & \nodata & \nodata & ... \\
          15 & A01300+1804 & HII & 1 & HII & 1 & HII \\
          16 & A01344+2838 & HII & 1 & HII & 1 & HII \\
          17 & A01346+0438 & HII & 1 & HII & 1 & HII \\
          18 & A01374+1539B & \nodata & \nodata & \nodata & \nodata & ... \\
          19 & NGC 695 & HII & 1 & HII & 1 & HII \\
          20 & NGC 784 & \nodata & \nodata & \nodata & \nodata & ... \\
          21 & A02008+2350 & HII & 1 & \nodata & \nodata & HII \\
          22 & IC 195 & \nodata & \nodata & \nodata & \nodata & ... \\
          23 & IC 197 & HII & 1 & \nodata & \nodata & HII \\
          24 & IC 1776 & HII & 1 & HII & 1 & HII \\
          25 & A02056+1444 & HII & 1 & HII & 1 & HII \\
          26 & NGC 825 & \nodata & \nodata & \nodata & \nodata & ... \\
          27 & NGC 927 & HII & 3 & \nodata & \nodata & \nodata \\
          28 & A02257-0134 & HII & 1 & HII & 1 & HII \\
          29 & NGC 984 & \nodata & \nodata & AGN & 1 & AGN \\
          30 & NGC 1029 & \nodata & \nodata & \nodata & \nodata & ... \\
          31 & A02464+1807 & \nodata & \nodata & \nodata & \nodata & ... \\
          32 & A02493-0122 & HII & 1 & \nodata & \nodata & HII \\
          33 & NGC 1298 & \nodata & \nodata & \nodata & \nodata & ... \\
          34 & A03202-0205 & HII & 1 & HII & 1 & HII \\
          35 & NGC 1552 & \nodata & \nodata & AGN & \nodata & AGN \\
          36 & NGC 2692 & \nodata & \nodata & \nodata & \nodata & ... \\
          37 & A08567+5242 & HII & 3 & AGN & \nodata & AGN \\
          38 & A09045+3328 & HII & 1 & HII & 1 & HII \\
          39 & NGC 2780 & HII & 3 & HII & 3 & HII \\
          40 & A09125+5303 & \nodata & \nodata & \nodata & \nodata & ... \\
          41 & NGC 2799 & HII & 1 & HII & 1 & HII \\
          42 & NGC 2824 & \nodata & \nodata & AGN & 1 & AGN \\
          43 & NGC 2844 & HII & 1 & HII & 1 & HII \\
          44 & NGC 3011 & AMB & 1 & AMB & 1 & AMB \\
          45 & NGC 3009 & HII & 1 & HII & 3 & HII \\
          46 & IC 2520 & HII & 1 & HII & 1 & HII \\
          47 & A09557+4758 & HII & 1 & HII & 1 & HII \\
          48 & NGC 3075 & HII & 1 & HII & 1 & HII \\
          49 & A09579+0439 & HII & 1 & HII & 3 & HII \\
          50 & NGC 3104 & HII & 1 & \nodata & \nodata & HII \\
          51 & A10042+4716 & HII & 1 & HII & 1 & HII \\
          52 & NGC 3165 & HII & 1 & AMB & 1 & AMB \\
          53 & A10114+0716 & HII & 1 & \nodata & \nodata & HII \\
          54 & NGC 3179 & \nodata & \nodata & \nodata & \nodata & ... \\
          55 & A10171+3853 & HII & 1 & HII & 1 & HII \\
          56 & NGC 3213 & HII & 3 & HII & 3 & HII \\
          57 & NGC 3264 & HII & 1 & HII & 1 & HII \\
          58 & NGC 3279 & HII & 3 & HII & 1 & HII \\
          59 & A10321+4649 & HII & 1 & HII & 1 & HII \\
          60 & A10337+1358 & HII & 1 & HII & 1 & HII \\
          61 & IC 2591 & HII & 1 & HII & 1 & HII \\
          62 & A10365+4812 & HII & 1 & HII & 1 & HII \\
          63 & A10368+4811 & HII & 1 & HII & 1 & HII \\
          64 & NGC 3326 & HII & 3 & HII & 1 & HII \\
          65 & A10389+3859 & \nodata & \nodata & \nodata & \nodata & ... \\
          66 & A10431+3514 & HII & 3 & \nodata & \nodata & HII \\
          67 & A10448+0731 & \nodata & \nodata & \nodata & \nodata & ... \\
          68 & A10465+0711 & HII & 1 & HII & 1 & HII \\
          69 & A10504+0454 & HII & 1 & HII & 1 & HII \\
          70 & NGC 3454 & HII & 1 & HII & 3 & HII \\  
          71 & A10592+1652 & HII & 1 & HII & 1 & HII \\
          72 & NGC 3499 & \nodata & \nodata & \nodata & \nodata & ... \\      \
          73 & NGC 3510 & HII & 1 & HII & 1 & HII \\
          74 & A11017+3828W & \nodata & \nodata & \nodata & \nodata & ... \\
          75 & NGC 3522 & \nodata & \nodata & \nodata & \nodata & ... \\
          76 & A11040+5130 & HII & 1 & HII & 3 & HII \\
          77 & IC 673 & HII & 1 & HII & 1 & HII \\
          78 & A11068+4705 & \nodata & \nodata & \nodata & \nodata & ... \\
          79 & A11072+1302 & HII & 1 & HII & 1 & HII \\
          80 & NGC 3605 & \nodata & \nodata & \nodata & \nodata & ... \\
          81 & A11142+1804 & HII & 3 & HII & 1 & HII \\
          82 & NGC 3633 & HII & 1 & HII & 1 & HII \\
          83 & IC 692 & AMB & 1 & AMB & 1 & AMB \\
          84 & A11238+5401 & \nodata & \nodata & \nodata & \nodata & ... \\
          85 & A11310+3254 & HII & 3 & HII & 1 & HII \\
          86 & IC 708 & \nodata & \nodata & \nodata & \nodata & ... \\
          87 & A11332+3536 & HII & 1 & HII & 1 & HII \\
          88 & A11336+5829 & HII & 1 & HII & 1 & HII \\
          89 & NGC 3795A & HII & 1 & HII & 3 & HII \\
          90 & A11372+2012 & HII & 1 & HII & 1 & HII \\
          91 & NGC 3795 & HII & 1 & HII & 3 & HII \\
          92 & A11378+2840 & HII & 1 & HII & 1 & HII \\
          93 & A11392+1615 & HII & 1 & HII & 1 & HII \\
          94 & NGC 3846 & HII & 1 & HII & 1 & HII \\
          95 & NGC 3850 & \nodata & \nodata & \nodata & \nodata & ... \\
          96 & A11476+4220 & HII & 1 & HII & 1 & HII \\
          97 & NGC 3913 & \nodata & \nodata & \nodata & \nodata & ... \\
          98 & IC 746 & HII & 1 & HII & 1 & HII \\
          99 & A11531+0132 & \nodata & \nodata & \nodata & \nodata & ... \\
         100 & NGC 3978 & HII & 1 & HII & 1 & HII \\
         101 & A11547+4933 & \nodata & \nodata & \nodata & \nodata & ... \\
         102 & A11547+5813 & HII & 1 & HII & 1 & HII \\
         103 & NGC 4034 & HII & 1 & \nodata & \nodata & HII \\
         104 & A11592+6237 & HII & 1 & HII & 1 & HII \\
         105 & A12001+6439 & HII & 1 & HII & 1 & HII \\
         106 & NGC 4117 & AMB & 1 & AMB & 1 & AMB \\
         107 & NGC 4120 & HII & 1 & HII & 1 & HII \\
         108 & A12064+4201 & HII & 3 & HII & 1 & HII \\
         109 & NGC 4141 & HII & 1 & HII & 1 & HII \\
         110 & NGC 4159 & HII & 1 & HII & 1 & HII \\
         111 & NGC 4204 & \nodata & \nodata & \nodata & \nodata & ... \\
         112 & NGC 4238 & HII & 1 & HII & 1 & HII \\
         113 & NGC 4248 & HII & 1 & HII & 1 & HII \\
         114 & A12167+4938 & HII & 1 & HII & 3 & HII \\
         115 & NGC 4272 & \nodata & \nodata & \nodata & \nodata & ... \\
         116 & NGC 4288 & HII & 1 & HII & 1 & HII \\
         117 & NGC 4308 & \nodata & \nodata & \nodata & \nodata & ... \\
         118 & A12195+3222 & \nodata & \nodata & \nodata & \nodata & ... \\
         119 & A12195+7535 & Sy1 & 1 & Sy1 & 1 & AGN \\
         120 & A12263+4331 & \nodata & \nodata & \nodata & \nodata & ... \\
         121 & A12295+4007 & HII & 1 & HII & 1 & HII \\
         122 & A12300+4259 & HII & 1 & \nodata & \nodata & HII \\
         123 & A12304+3754 & HII & 1 & HII & 1 & HII \\
         124 & NGC 4509 & HII & 1 & HII & 1 & HII \\
         125 & A12331+7230 & HII & 1 & HII & 1 & HII \\
         126 & A12446+5155 & HII & 1 & HII & 1 & HII \\
         127 & NGC 4758 & HII & 1 & HII & 1 & HII \\
         128 & NGC 4795 & \nodata & \nodata & \nodata & \nodata & ... \\
         129 & NGC 4807 & \nodata & \nodata & \nodata & \nodata & ... \\
         130 & NGC 4841B & \nodata & \nodata & \nodata & \nodata & ... \\
         131 & NGC 4926 & \nodata & \nodata & \nodata & \nodata & ... \\
         132 & NGC 4961 & \nodata & \nodata & \nodata & \nodata & ... \\
         133 & A13065+5420 & HII & 1 & HII & 1 & HII \\
         134 & IC 4213 & HII & 1 & HII & 1 & HII \\
         135 & A13194+4232 & HII & 1 & HII & 3 & HII \\
         136 & NGC 5117 & \nodata & \nodata & \nodata & \nodata & ... \\
         137 & NGC 5173 & AMB & 1 & AMB &        1 & AMB \\
         138 & A13281+3153 & \nodata & \nodata & \nodata & \nodata & ... \\
         139 & NGC 5208 & \nodata & \nodata & \nodata & \nodata & ... \\
         140 & NGC 5230 & HII & 1 & HII & 3 & HII \\
         141 & A13361+3323 & HII & 1 & HII & 1 & HII \\
         142 & NGC 5267 & \nodata & \nodata & \nodata & \nodata & ... \\
         143 & A13422+3526 & HII & 1 & HII & 1 & HII \\
         144 & NGC 5338 & HII & 3 & HII & 1 & HII \\
         145 & NGC 5356 & HII & 3 & HII & 3 & HII \\
         146 & A13550+4613 & \nodata & \nodata & \nodata & \nodata & ... \\
         147 & NGC 5407 & \nodata & \nodata & \nodata & \nodata & ... \\
         148 & NGC 5425 & HII & 1 & HII & 1 & HII \\
         149 & A14016+3559 & \nodata & \nodata & \nodata & \nodata & ... \\
         150 & NGC 5470 & \nodata & \nodata & \nodata & \nodata & ... \\
         151 & NGC 5491 & HII & 3 & \nodata & \nodata & HII \\
         152 & NGC 5532 & \nodata & \nodata & \nodata & \nodata & ... \\
         153 & NGC 5541 & HII & 1 & \nodata & \nodata & HII \\
         154 & NGC 5596 & \nodata & \nodata & \nodata & \nodata & ... \\
         155 & NGC 5608 & HII & 1 & HII & 1 & HII \\
         156 & A14305+1149 & HII & 1 & HII & 1 & HII \\
         157 & NGC 5684 & \nodata & \nodata & AGN & \nodata & AGN \\
         158 & NGC 5762 & HII & 1 & \nodata & \nodata & HII \\
         159 & A14489+3547 & HII & 1 & HII & 1 & HII \\
         160 & A14492+3545 & HII & 1 & HII & 1 & HII \\
         161 & IC 1066 & HII & 1 & \nodata & \nodata & HII \\
         162 & A14594+4454 & AMB & 1 & AMB & 1 & AMB \\
         163 & A15016+1037 & Sy1 & 1 & Sy1 & 1 & AGN \\
         164 & IC 1100 & HII & 3 & HII & 3 & HII \\
         165 & NGC 5874 & HII & 1 & HII & 3 & HII \\
         166 & NGC 5875A & HII & 1 & HII & 1 & HII \\
         167 & NGC 5888 & \nodata & \nodata & \nodata & \nodata & ... \\
         168 & IC 1124 & HII & 1 & HII & 3 & HII \\
         169 & NGC 5940 & Sy1 & 1 & Sy1 & 1 & AGN \\
         170 & A15314+6744 & HII & 1 & HII & 3 & HII \\
         171 & NGC 5993 & HII & 1 & HII & 1 & HII \\
         172 & IC 1141 & AMB & 1 & AMB & 1 & AMB \\
         173 & IC 1144 & \nodata & \nodata & \nodata & \nodata & ... \\
         174 & NGC 6007 & HII & 1 & \nodata & \nodata & HII \\
         175 & A15523+1645 & HII & 1 & HII & 1 & HII \\
         176 & A15542+4800 & \nodata & \nodata & \nodata & \nodata & ... \\
         177 & NGC 6020 & \nodata & \nodata & \nodata & \nodata & ... \\
         178 & NGC 6123 & \nodata & \nodata & \nodata & \nodata & ... \\
         179 & NGC 6131 & HII & 1 & HII & 3 & HII \\
         180 & NGC 6185 & \nodata & \nodata & \nodata & \nodata & ... \\
         181 & NGC 7077 & HII & 1 & HII & 1 & HII \\
         182 & NGC 7194 & \nodata & \nodata & \nodata & \nodata & ... \\
         183 & A22306+0750 & HII & 1 & HII & 1 & HII \\
         184 & NGC 7328 & HII & 3 & HII & 3 & HII \\
         185 & NGC 7360 & \nodata & \nodata & AMB & 1 & AMB \\
         186 & A22426+0610 & HII & 1 & \nodata & \nodata & HII \\
         187 & A22551+1931N & HII & 1 & HII & 1 & HII \\
         188 & NGC 7436 & \nodata & \nodata & \nodata & \nodata & ... \\
         189 & NGC 7460 & HII & 1 & HII & 1 & HII \\
         190 & NGC 7537 & HII & 1 & HII & 1 & HII \\
         191 & NGC 7548 & \nodata & \nodata & \nodata & \nodata & ... \\
         192 & A23176+1541 & HII & 1 & HII & 3 & HII \\
         193 & NGC 7620 & HII & 1 & HII & 3 & HII \\
         194 & A23264+1703 & \nodata & \nodata & \nodata & \nodata & ... \\
         195 & IC 1504 & HII & 2 & HII & 3 & HII \\
         196 & NGC 7752 & HII & 1 & HII & 1 & HII \\
         197 & A23514+2813 & \nodata & \nodata & \nodata & \nodata & ... \\
         198 & A23542+1633 & HII & 1 & HII & 1 & HII \\
\enddata
\tablenotetext{a}{Classes 'HII', 'AMB' and 'AGN' are defined according to the  \citet{Kewley01b} theoretical classification scheme.}
\tablenotetext{b}{Quality flags, $Q$, are defined as: 1= object classified using [NII]/H$\alpha$, [SII]/H$\alpha$, and [OIII]/H$\beta$; 2= object classified using [NII]/H$\alpha$ and [OIII]/H$\beta$ only; 3= object classified using [NII]/H$\alpha$ only.}
\end{deluxetable}

\newpage

\LongTables
\begin{deluxetable}{rlrrlllllccrrll}
\setlength{\tabcolsep}{5.5pt}
\tablecolumns{15}
\tablewidth{\linewidth}
\tabletypesize{\scriptsize}
\tablecaption{Properties of the 101 galaxies selected from the NFGS
\label{sample_table}}
\tablehead{
  \multicolumn{1}{c}{ID}             &
  \multicolumn{1}{c}{Name}           &
  \multicolumn{1}{c}{T}              &
  \multicolumn{1}{c}{$cz$}           &
  \multicolumn{1}{c}{$M_B$}          &
  \multicolumn{1}{c}{$\mu_0^B$}      &
  \multicolumn{2}{c}{$I/I_{B{26}}$}  &
  \multicolumn{1}{c}{$\zeta$}        &
  \multicolumn{2}{c}{log(O/H)+12}    &
  \multicolumn{2}{c}{$A(V)$}         &
  \multicolumn{2}{c}{SFR(H$\alpha$)} \\
  &    &    &    &    &    &
  \multicolumn{1}{c}{nuc}  &
  \multicolumn{1}{c}{int}  &
  &                           
  \multicolumn{1}{c}{nuc}  &
  \multicolumn{1}{c}{int}  &
  \multicolumn{1}{c}{nuc}  &
  \multicolumn{1}{c}{int}  &
  \multicolumn{1}{c}{nuc}  &
  \multicolumn{1}{c}{int}  \\
  \multicolumn{1}{c}{(1)}  &
  \multicolumn{1}{c}{(2)}  &
  \multicolumn{1}{c}{(3)}  &
  \multicolumn{1}{c}{(4)}  &
  \multicolumn{1}{c}{(5)}  &
  \multicolumn{1}{c}{(6)}  &
  \multicolumn{1}{c}{(7)}  &
  \multicolumn{1}{c}{(8)}  &
  \multicolumn{1}{c}{(9)}  &
  \multicolumn{1}{c}{(10)} &
  \multicolumn{1}{c}{(11)} &
  \multicolumn{1}{c}{(12)} &
  \multicolumn{1}{c}{(13)} &
  \multicolumn{1}{c}{(14)} &
  \multicolumn{1}{c}{(15)}  
}
\startdata
  2 & A00289+0556  &  7 &  2055 &--17.84 & 20.64 & 0.058 & 0.919 & 0.063 & 8.82 & 8.71 & $<0.02$ & 0.93 & 0.006  & 0.259 \\
  4 & A00389-0159  &  1 &  5302 &--20.52 & 18.89 & 0.098 & 0.862 & 0.113 & 8.94 & 8.94 & 2.22 & 1.46 & 0.394  & 4.985 \\
  5 & A00442+3224  &  3 &  4859 &--20.31 & 20.17 & 0.039 & 0.771 & 0.051 & 9.08 & 8.92 & 1.99 & 1.23 & 0.286  & 3.191 \\
 15 & A01300+1804  & 10 &   686 &--15.90 & 20.46 & 0.058 & 0.869 & 0.067 & 8.06 & 8.06 & 0.50 & 0.50 & 0.001  & 0.022 \\
 16 & A01344+2838  &  4 &  7756 &--20.53 & 19.35 & 0.142 & 0.904 & 0.157 & 8.84 & 8.78 & 1.79 & 1.41 & 0.416  & 6.929 \\
 17 & A01346+0438  &  4 &  3158 &--18.40 & 19.85 & 0.188 & 0.957 & 0.197 & 8.85 & 8.77 & 0.81 & 0.98 & 0.142  & 0.802 \\
 19 & NGC695       &  5 &  9705 &--21.71 & 20.12 & 0.081 & 0.832 & 0.098 & 8.96 & 8.89 & 2.72 & 2.49 & 9.034  & 104.2 \\
 23 & IC197        &  4 &  6332 &--20.20 & 20.30 & 0.073 & 0.854 & 0.086 & 8.97 & 8.87 & 1.96 & 1.41 & 0.212  & 4.573 \\
 24 & IC1776       &  5 &  3405 &--19.27 & 21.58 & 0.019 & 0.903 & 0.021 & 8.77 & 8.54 & 0.04 & 0.62 & 0.008  & 0.908 \\
 25 & A02056+1444  &  3 &  4405 &--19.99 & 20.33 & 0.048 & 0.715 & 0.067 & 8.90 & 8.90 & 2.35 & 1.46 & 0.412  & 3.008 \\
 28 & A02257-0134  &  8 &  1762 &--17.76 & 21.66 & 0.020 & 0.522 & 0.038 & 9.05 & 8.72 & 0.33 & 0.78 & 0.002  & 0.119 \\
 34 & A03202-0205  &  1 &  8227 &--20.92 & 19.68 & 0.099 & 0.860 & 0.115 & 9.10 & 8.93 & 1.13 & 1.35 & 0.330  & 5.445 \\
 38 & A09045+3328  &  8 &   553 &--15.20 & 21.68 & 0.025 & 0.662 & 0.037 & 7.94 & 8.67 & $<0.02$ & 0.13 & 0.0001 & 0.005 \\
 39 & NGC2780      &  2 &  1951 &--17.91 & 21.05 & 0.037 & 0.795 & 0.046 & 9.11 & 9.03 & 1.20 & 1.12 & 0.024  & 0.154 \\
 41 & NGC2799      &  9 &  1882 &--18.13 & 20.11 & 0.094 & 0.815 & 0.115 & 8.96 & 8.84 & 0.47 & 0.97 & 0.050  & 0.315 \\
 43 & NGC2844      &  1 &  1486 &--18.18 & 19.17 & 0.091 & 0.864 & 0.106 & 8.68 & 8.81 & 2.47 & 1.22 & 0.026  & 0.166 \\
 45 & NGC3009      &  5 &  4666 &--19.50 & 19.92 & 0.073 & 0.787 & 0.092 & 9.23 & 8.93 & 1.43 & 1.00 & 0.100  & 0.828 \\
 46 & IC2520       &... &  1226 &--17.53 & 19.88 & 0.068 & 0.719 & 0.094 & 8.91 & 8.84 & 1.42 & 1.51 & 0.028  & 0.574 \\
 47 & A09557+4758  &  9 &  1172 &--17.90 & 21.40 & 0.018 & 0.780 & 0.024 & 7.90 & 8.51 & 0.17 & 0.12 & 0.001  & 0.125 \\
 48 & NGC3075      &  5 &  3566 &--19.49 & 20.09 & 0.055 & 0.794 & 0.069 & 9.13 & 8.97 & 0.96 & 0.49 & 0.175  & 0.795 \\
 49 & A09579+0439  &  3 &  4185 &--19.14 & 20.75 & 0.072 & 0.918 & 0.078 & 9.01 & 8.76 & 1.29 & 1.52 & 0.065  & 1.491 \\
 51 & A10042+4716  & 10 &   571 &--15.88 & 21.85 & 0.017 & 0.762 & 0.022 & 8.72 & 8.67 & 0.02 & 0.18 & 0.0001 & 0.012 \\
 55 & A10171+3853  &  9 &  2008 &--18.10 & 21.04 & 0.034 & 0.761 & 0.044 & 8.79 & 8.85 & 0.54 & 0.27 & 0.004  & 0.130 \\
 56 & NGC3213      &  4 &  1412 &--17.90 & 20.58 & 0.035 & 0.791 & 0.045 & 8.88 & 8.81 & $<0.02$ & 1.32 & 0.0004 & 0.120 \\
 57 & NGC3264      &  8 &   929 &--17.58 & 21.41 & 0.016 & 0.788 & 0.020 & 8.65 & 8.62 & $<0.02$ & $<0.02$ & 0.002  & 0.118 \\
 58 & NGC3279      &  5 &  1422 &--18.04 & 20.92 & 0.020 & 0.866 & 0.023 & 9.02 & 8.89 & 1.24 & 1.81 & 0.005  & 0.300 \\
 59 & A10321+4649  &  5 &  3338 &--18.91 & 19.31 & 0.150 & 0.782 & 0.192 & 9.07 & 9.06 & 1.09 & 0.97 & 0.181  & 0.717 \\
 60 & A10337+1358  &  6 &  2997 &--18.83 & 20.92 & 0.035 & 0.756 & 0.046 & 8.91 & 8.89 & 2.03 & 1.43 & 0.050  & 0.974 \\
 61 & IC2591       &  4 &  6755 &--20.49 & 19.62 & 0.107 & 0.877 & 0.122 & 8.98 & 8.79 & 0.88 & 0.67 & 0.290  & 3.077 \\
 62 & A10365+4812  &  5 &   854 &--16.34 & 20.07 & 0.089 & 0.713 & 0.124 & 8.14 & 8.06 & 0.10 & $<0.02$ & 0.002  & 0.018 \\
 63 & A10368+4811  &  5 &  1534 &--17.41 & 20.86 & 0.066 & 0.794 & 0.083 & 8.22 & 8.35 & 0.49 & 0.41 & 0.004  & 0.090 \\
 64 & NGC3326      &  3 &  8136 &--20.78 & 18.65 & 0.215 & 0.815 & 0.263 & 9.09 & 8.98 & 2.44 & 1.89 & 3.126  & 8.130 \\
 68 & A10465+0711  &  0 &   722 &--14.92 & 20.10 & 0.101 & 0.776 & 0.130 & 8.63 & 8.26 & 0.37 & 0.44 & 0.002  & 0.012 \\
 69 & A10504+0454  &  0 &  5793 &--19.97 & 18.14 & 0.334 & 0.882 & 0.378 & 8.92 & 8.91 & 0.91 & 0.73 & 1.990  & 2.617 \\
 70 & NGC3454      &  5 &  1153 &--17.37 & 21.50 & 0.020 & 0.843 & 0.024 & 8.81 & 8.79 & 1.43 & 0.76 & 0.003  & 0.113 \\
 71 & A10592+1652  &  4 &  2936 &--18.44 & 21.29 & 0.045 & 0.738 & 0.062 & 8.77 & 8.57 & 0.06 & 0.48 & 0.007  & 0.241 \\
 73 & NGC3510      &  7 &   704 &--16.39 & 20.69 & 0.028 & 0.859 & 0.032 & 8.65 & 8.59 & 0.06 & 0.15 & 0.001  & 0.081 \\
 76 & A11040+5130  &  5 &  2204 &--19.25 & 21.13 & 0.014 & 0.677 & 0.021 & 8.79 & 8.74 & $<0.02$ & 0.20 & 0.002  & 0.276 \\
 77 & IC673        &  1 &  3851 &--19.45 & 20.14 & 0.079 & 0.815 & 0.097 & 9.05 & 8.96 & 1.76 & 1.24 & 0.163  & 1.355 \\
 79 & A11072+1302  &  5 & 12743 &--21.38 & 20.29 & 0.133 & 0.898 & 0.148 & 8.93 & 8.79 & 1.15 & 1.00 & 1.937  & 16.24 \\
 81 & A11142+1804  &  5 &   973 &--16.04 & 21.32 & 0.033 & 0.763 & 0.044 & 9.03 & 8.87 & 1.44 & 1.80 & 0.003  & 0.074 \\
 82 & NGC3633      &  1 &  2553 &--18.92 & 19.66 & 0.109 & 0.877 & 0.124 & 9.15 & 8.91 & 2.34 & 2.73 & 0.252  & 2.226 \\
 85 & A11310+3254  &  3 &  2619 &--18.69 & 20.36 & 0.059 & 0.715 & 0.082 & 9.03 & 8.69 & 1.38 & 1.79 & 0.012  & 0.253 \\
 87 & A11332+3536  &--3 &  1598 &--17.89 & 18.75 & 0.190 & 0.847 & 0.225 & 8.82 & 8.78 & 1.33 & 0.80 & 0.106  & 0.134 \\
 88 & A11336+5829  &  5 &  1225 &--17.45 & 21.28 & 0.032 & 0.541 & 0.058 & 8.34 & 8.54 & $<0.02$ & 0.37 & 0.002  & 0.049 \\
 89 & NGC3795A     &  6 &  1154 &--18.03 & 21.89 & 0.010 & 0.780 & 0.012 & 8.75 & 8.62 & 2.88 & $<0.02$ & 0.001  & 0.059 \\
 90 & A11372+2012  &  5 & 10964 &--21.60 & 19.79 & 0.097 & 0.875 & 0.111 & 9.27 & 8.92 & 1.29 & 0.92 & 0.480  & 13.13 \\
 91 & NGC3795      &  5 &  1091 &--17.48 & 20.71 & 0.020 & 0.796 & 0.025 & 8.95 & 8.72 & 0.52 & 0.90 & 0.0005 & 0.057 \\
 92 & A11378+2840  &--3 &  1821 &--17.59 & 19.94 & 0.164 & 0.886 & 0.184 & 8.65 & 8.61 & 0.33 & 0.49 & 0.015  & 0.104 \\
 93 & A11392+1615  &--2 &   786 &--14.82 & 19.80 & 0.171 & 0.752 & 0.228 & 8.63 & 8.62 & 0.17 & 0.18 & 0.009  & 0.026 \\
 94 & NGC3846      &  9 &  1396 &--18.22 & 20.92 & 0.023 & 0.825 & 0.028 & 8.71 & 8.66 & 0.20 & $<0.02$ & 0.002  & 0.135 \\
 96 & A11476+4220  &--2 &  1033 &--16.62 & 19.26 & 0.254 & 0.794 & 0.320 & 8.94 & 8.91 & 0.54 & 0.74 & 0.030  & 0.068 \\
 98 & IC746        &  3 &  5027 &--19.59 & 20.11 & 0.103 & 0.830 & 0.124 & 8.89 & 8.66 & 0.53 & 0.98 & 0.065  & 1.458 \\
100 & NGC3978      &  4 &  9978 &--22.22 & 20.23 & 0.037 & 0.877 & 0.043 & 9.29 & 9.02 & 1.17 & 1.28 & 0.447  & 25.74 \\
102 & A11547+5813  &  9 &  1175 &--17.06 & 21.91 & 0.025 & 0.842 & 0.029 & 8.19 & 8.17 & 0.72 & 0.22 & 0.001  & 0.042 \\
104 & A11592+6237  & 10 &  1120 &--17.08 & 21.44 & 0.033 & 0.828 & 0.039 & 8.37 & 8.27 & $<0.02$ & $<0.02$ & 0.005  & 0.049 \\
105 & A12001+6439  &--2 &  1447 &--17.62 & 18.50 & 0.242 & 0.906 & 0.268 & 8.89 & 8.83 & 1.00 & 1.02 & 0.157  & 0.409 \\
107 & NGC4120      &  5 &  2251 &--18.61 & 21.12 & 0.038 & 0.748 & 0.050 & 8.71 & 8.69 & 0.69 & 0.46 & 0.015  & 0.332 \\
108 & A12064+4201  &  2 &   927 &--17.05 & 20.32 & 0.063 & 0.790 & 0.080 & 9.09 & 8.98 & 1.31 & 1.37 & 0.019  & 0.076 \\
109 & NGC4141      &  5 &  1980 &--19.08 & 20.54 & 0.035 & 0.857 & 0.041 & 8.74 & 8.60 & $<0.02$ & 0.23 & 0.007  & 0.565 \\
110 & NGC4159      &  8 &  1761 &--18.13 & 20.63 & 0.047 & 0.866 & 0.054 & 8.72 & 8.64 & 0.66 & 0.94 & 0.010  & 0.369 \\
112 & NGC4238      &  5 &  2771 &--18.96 & 21.10 & 0.036 & 0.782 & 0.046 & 8.87 & 8.75 & 0.42 & 0.45 & 0.034  & 0.586 \\
113 & NGC4248      &  8 &   484 &--16.31 & 21.32 & 0.013 & 0.808 & 0.015 & 8.27 & 8.70 & 2.28 & 1.01 & 0.001  & 0.023 \\
114 & A12167+4938  &  5 &  3639 &--19.38 & 20.85 & 0.036 & 0.801 & 0.045 & 9.05 & 8.74 & 0.64 & 1.07 & 0.023  & 0.912 \\
116 & NGC4288      &  9 &   532 &--16.40 & 20.46 & 0.023 & 0.839 & 0.028 & 8.71 & 8.52 & $<0.02$ & 0.41 & 0.001  & 0.044 \\
121 & A12295+4007  &  7 &   685 &--16.04 & 20.97 & 0.037 & 0.885 & 0.042 & 8.25 & 7.93 & $<0.02$ & $<0.02$ & 0.001  & 0.021 \\
123 & A12304+3754  &  7 &   503 &--16.07 & 21.02 & 0.014 & 0.791 & 0.018 & 8.64 & 8.65 & $<0.02$ & 0.07 & 0.0004 & 0.023 \\
124 & NGC4509      &  9 &   907 &--16.73 & 20.57 & 0.066 & 0.845 & 0.078 & 8.25 & 8.21 & $<0.02$ & 0.11 & 0.019  & 0.095 \\
125 & A12331+7230  &  3 &  6959 &--20.38 & 20.29 & 0.075 & 0.773 & 0.097 & 8.99 & 8.85 & 1.09 & 1.43 & 0.298  & 5.127 \\
126 & A12446+5155  & 10 &   502 &--16.48 & 20.75 & 0.028 & 0.786 & 0.035 & 8.04 & 7.88 & $<0.02$ & $<0.02$ & 0.001  & 0.012 \\
127 & NGC4758      &  4 &  1244 &--18.35 & 21.46 & 0.011 & 0.762 & 0.014 & 8.30 & 8.62 & 2.56 & 1.29 & 0.008  & 0.225 \\
133 & A13065+5420  &  3 &  2460 &--18.46 & 21.10 & 0.040 & 0.798 & 0.050 & 8.74 & 8.52 & 0.34 & 0.21 & 0.008  & 0.308 \\
134 & IC4213       &  6 &   815 &--16.74 & 21.43 & 0.022 & 0.834 & 0.026 & 8.63 & 8.57 & $<0.02$ & 0.38 & 0.001  & 0.041 \\
135 & A13194+4232  &  6 &  3396 &--19.03 & 20.48 & 0.056 & 0.800 & 0.070 & 8.95 & 8.83 & 0.84 & 0.88 & 0.039  & 0.643 \\
141 & A13361+3323  &  9 &  2364 &--18.53 & 21.52 & 0.019 & 0.668 & 0.029 & 8.61 & 8.36 & 0.26 & 0.53 & 0.011  & 0.331 \\
143 & A13422+3526  &  4 &  2502 &--18.38 & 20.74 & 0.060 & 0.784 & 0.076 & 8.81 & 8.70 & 1.03 & 1.02 & 0.013  & 0.226 \\
145 & NGC5356      &  3 &  1397 &--18.71 & 21.02 & 0.016 & 0.760 & 0.021 & 9.08 & 8.67 & 1.05 & 3.29 & 0.002  & 0.929 \\
148 & NGC5425      &  5 &  2062 &--18.48 & 21.00 & 0.035 & 0.792 & 0.044 & 8.93 & 8.76 & 0.97 & 0.78 & 0.008  & 0.276 \\
153 & NGC5541      &  5 &  7698 &--21.29 & 19.86 & 0.071 & 0.903 & 0.079 & 8.96 & 8.94 & 1.22 & 1.47 & 0.253  & 12.53 \\
155 & NGC5608      &  9 &   662 &--16.48 & 22.04 & 0.015 & 0.861 & 0.017 & 7.96 & 8.03 & $<0.02$ & $<0.02$ & 0.0001 & 0.014 \\
156 & A14305+1149  &  5 &  2234 &--18.61 & 20.84 & 0.036 & 0.898 & 0.040 & 8.90 & 8.77 & 0.53 & 0.39 & 0.005  & 0.268 \\
159 & A14489+3547  &... &  1215 &--16.94 & 18.85 & 0.323 & 0.866 & 0.373 & 8.26 & 8.24 & $<0.02$ & $<0.02$ & 0.069  & 0.167 \\
160 & A14492+3545  &... &  1306 &--17.35 & 20.08 & 0.092 & 0.863 & 0.106 & 8.55 & 8.24 & 0.09 & 0.28 & 0.005  & 0.066 \\
161 & IC1066       &  2 &  1613 &--18.31 & 20.01 & 0.061 & 0.907 & 0.068 & 8.95 & 8.72 & 0.46 & 1.53 & 0.001  & 0.308 \\
164 & IC1100       &  6 &  6561 &--20.71 & 19.67 & 0.086 & 0.834 & 0.103 & 9.17 & 9.00 & 0.83 & 1.57 & 0.154  & 4.428 \\
165 & NGC5874      &  4 &  3128 &--19.83 & 21.24 & 0.014 & 0.792 & 0.018 & 8.79 & 8.80 & $<0.02$ & 1.60 & 0.002  & 1.066 \\
166 & NGC5875A     &  5 &  2470 &--18.52 & 20.47 & 0.068 & 0.642 & 0.106 & 8.91 & 8.92 & 0.57 & 0.40 & 0.028  & 0.256 \\
168 & IC1124       &  2 &  5242 &--19.86 & 19.79 & 0.094 & 0.871 & 0.108 & 8.93 & 8.97 & 2.69 & 1.99 & 0.347  & 3.223 \\
170 & A15314+6744  &  5 &  6461 &--20.49 & 20.54 & 0.037 & 0.868 & 0.042 & 9.30 & 8.99 & 0.51 & 0.70 & 0.028  & 1.841 \\
171 & NGC5993      &  3 &  9578 &--21.65 & 19.88 & 0.060 & 0.826 & 0.073 & 9.09 & 9.02 & 0.72 & 0.83 & 0.198  & 8.109 \\
175 & A15523+1645  &  5 &  2191 &--17.84 & 20.12 & 0.142 & 0.820 & 0.173 & 8.79 & 8.62 & 0.27 & 0.84 & 0.029  & 0.289 \\
179 & NGC6131      &  5 &  5054 &--20.46 & 20.97 & 0.029 & 0.795 & 0.036 & 9.11 & 8.99 & 0.92 & 0.41 & 0.076  & 1.113 \\
181 & NGC7077      &  0 &  1142 &--16.88 & 19.38 & 0.167 & 0.823 & 0.203 & 8.65 & 8.54 & 0.34 & 0.48 & 0.016  & 0.097 \\
183 & A22306+0750  &  5 &  1995 &--18.05 & 20.17 & 0.082 & 0.801 & 0.102 & 8.91 & 8.87 & 2.13 & 1.75 & 0.077  & 0.864 \\
187 & A22551+1931N &--2 &  5682 &--18.84 & 19.93 & 0.264 & 0.740 & 0.357 & 8.90 & 8.90 & 1.82 & 1.90 & 2.002  & 4.014 \\
189 & NGC7460      &  3 &  3296 &--19.92 & 20.01 & 0.031 & 0.728 & 0.042 & 9.17 & 8.97 & 1.99 & 1.47 & 0.263  & 3.124 \\
190 & NGC7537      &  4 &  2648 &--19.10 & 19.77 & 0.059 & 0.819 & 0.072 & 9.04 & 8.93 & 0.66 & 1.28 & 0.037  & 1.874 \\
193 & NGC7620      &  6 &  9565 &--21.87 & 18.75 & 0.108 & 0.913 & 0.118 & 8.96 & 8.88 & 2.12 & 1.74 & 2.280  & 38.98 \\
195 & IC1504       &  3 &  6306 &--20.56 & 20.67 & 0.041 & 0.797 & 0.051 & 9.18 & 8.78 & 2.47 & 2.84 & 0.464  & 14.86 \\
196 & NGC7752      &  7 &  4902 &--19.70 & 19.90 & 0.130 & 0.730 & 0.179 & 8.95 & 8.97 & 1.61 & 1.14 & 2.015  & 5.433 \\
198 & A23542+1633  & 10 &  1788 &--17.83 & 21.43 & 0.022 & 0.813 & 0.027 & 8.71 & 8.08 & 0.09 & 0.35 & 0.005  & 0.282
\enddata
\end{deluxetable}

\newpage

\begin{deluxetable}{llllllllllll}
\tablecolumns{11}
\tablewidth{\linewidth}
\tabletypesize{\scriptsize}
\tablecaption{Relative $B_{26}$ covering fraction $\zeta$, metallicity ratio, A(V) and SFR 
as a function of galaxy type.
\label{Type_properties}}
\tablehead{Hubble & \# & $z$ & Diam 
& \multicolumn{2}{c}{$\zeta$}
& \multicolumn{2}{c}{$\Delta$log(O/H)}
& \multicolumn{2}{c}{$\Delta$A(V)}
& \multicolumn{2}{c}{$\Delta$SFR}\\
Type &  & mean\tablenotemark{a} & (kpc) & mean\tablenotemark{a} & rms & mean\tablenotemark{a}\tablenotemark{b} & rms\tablenotemark{c} & mean\tablenotemark{a}\tablenotemark{d} & rms\tablenotemark{e} & mean\tablenotemark{a} & rms\\
}
\startdata
S0-Sab  & 18 & 0.0101 & 13.38 & $0.18\pm0.02$ & 0.10 & $0.09\pm0.03$ & 0.12 & 
               $0.13\pm 0.13$ & 0.50 & $1.47\pm0.23$ & 0.94  \\
Sb-Sd  & 61 & 0.0139 & 19.87      & $0.07\pm0.01$ & 0.05 & $0.14\pm0.02$ & 0.13 &  
               $0.07\pm0.10$ & 0.69 & $0.85\pm0.08$ & 0.63 \\
Sdm-Im-Pec & 22 & 0.0048 & 8.38 & $0.06\pm0.01$ & 0.08 & $0.11\pm0.06$ & 0.30 & 
               $-0.001\pm0.09$ &0.36 &  $0.96\pm0.13$ & 0.61 \\
\enddata
\tablenotetext{a}{The error quoted is the standard error of the mean}
\tablenotetext{b}{The mean of $\Delta$log(O/H) is the logarithm of the arithmetic mean of the (O/H) values}
\tablenotetext{c}{The rms of $\Delta$log(O/H) is calculated in log(O/H) units using the mean in {\it b}}
\tablenotetext{d}{The mean of $\Delta$A(V) is calculated by taking the mean Balmer Decrement and 
converting to A(V) assuming case b recombination and ${\rm R_{V} = A_{V}/E(B-V)}=3.1$}
\tablenotetext{e}{The rms of $\Delta$log(O/H) is calculated in A(V) units using the mean in {\it d}}
\end{deluxetable}

\begin{deluxetable}{rrcccc}
\tablecolumns{6}
\tablewidth{\linewidth}
\tabletypesize{\scriptsize}
\tablecaption{Mean elliptical radii at various relative and absolute fractions
	of the $B_{26}$ light \label{radius}}
\tablehead{ $\zeta$ & ${\rm I}_{B_{26}}$ & \multicolumn{4}{c}{mean elliptical radius (arcsec)} \\
        &             &         All         &    S0-Sab          &        Sb-Sd       &      Sdm-Im/Pec    \\
        &             & mean$\pm$error(rms) & mean$\pm$error(rms) & mean$\pm$error(rms) & mean$\pm$error(rms)\\
}
\startdata
   10\% &   8.1\% &  3.51$\pm$0.02( 1.68) &  2.04$\pm$0.01( 0.85) &  3.64$\pm$0.02( 1.63) &  4.34$\pm$0.02( 1.63) \\
   \nodata  &  10.0\% &  3.99$\pm$0.02( 1.92) &  2.33$\pm$0.01( 0.99) &  4.15$\pm$0.02( 1.88) &  4.92$\pm$0.02( 1.86) \\
   20\% &  16.2\% &  5.44$\pm$0.03( 2.64) &  3.23$\pm$0.01( 1.38) &  5.65$\pm$0.03( 2.60) &  6.67$\pm$0.04( 2.57) \\
   \nodata &  20.0\% &  6.25$\pm$0.03( 3.04) &  3.75$\pm$0.02( 1.59) &  6.49$\pm$0.03( 3.00) &  7.67$\pm$0.04( 2.95) \\
   30\% &  24.3\% &  7.15$\pm$0.04( 3.48) &  4.31$\pm$0.02( 1.82) &  7.39$\pm$0.04( 3.43) &  8.78$\pm$0.05( 3.39) \\
   \nodata  &  30.0\% &  8.31$\pm$0.05( 4.04) &  5.07$\pm$0.02( 2.10) &  8.57$\pm$0.05( 3.98) & 10.25$\pm$0.07( 3.97) \\
   40\% &  32.4\% &  8.80$\pm$0.05( 4.27) &  5.39$\pm$0.02( 2.22) &  9.06$\pm$0.05( 4.21) & 10.87$\pm$0.07( 4.21) \\
   \nodata  &  40.0\% & 10.36$\pm$0.07( 4.99) &  6.48$\pm$0.03( 2.60) & 10.61$\pm$0.07( 4.90) & 12.86$\pm$0.10( 4.98) \\
   50\% &  40.5\% & 10.47$\pm$0.07( 5.04) &  6.55$\pm$0.03( 2.62) & 10.71$\pm$0.07( 4.95) & 12.99$\pm$0.10( 5.03) \\
   60\% &  48.6\% & 12.22$\pm$0.09( 5.80) &  7.87$\pm$0.05( 3.08) & 12.43$\pm$0.09( 5.67) & 15.21$\pm$0.13( 5.88) \\
   \nodata  &  50.0\% & 12.54$\pm$0.09( 5.93) &  8.11$\pm$0.05( 3.16) & 12.74$\pm$0.09( 5.80) & 15.61$\pm$0.14( 6.04) \\
   70\% &  56.7\% & 14.15$\pm$0.12( 6.60) &  9.42$\pm$0.07( 3.64) & 14.28$\pm$0.12( 6.43) & 17.64$\pm$0.19( 6.84) \\
   \nodata  &  60.0\% & 15.00$\pm$0.14( 6.96) & 10.14$\pm$0.08( 3.91) & 15.09$\pm$0.13( 6.76) & 18.73$\pm$0.22( 7.30) \\
   80\% &  64.8\% & 16.32$\pm$0.17( 7.51) & 11.32$\pm$0.10( 4.39) & 16.33$\pm$0.16( 7.25) & 20.39$\pm$0.27( 8.01) \\
   \nodata  &  70.0\% & 17.88$\pm$0.22( 8.14) & 12.76$\pm$0.13( 4.99) & 17.79$\pm$0.20( 7.80) & 22.33$\pm$0.34( 8.84) \\
   90\% &  72.9\% & 18.83$\pm$0.25( 8.50) & 13.63$\pm$0.15( 5.31) & 18.69$\pm$0.23( 8.12) & 23.49$\pm$0.39( 9.33) \\
   \nodata  &  80.0\% & 21.51$\pm$0.37( 9.45) & 16.18$\pm$0.25( 6.22) & 21.19$\pm$0.34( 8.93) & 26.73$\pm$0.56(10.64) \\
  100\% &  81.0\% & 21.94$\pm$0.40( 9.60) & 16.61$\pm$0.27( 6.38) & 21.60$\pm$0.37( 9.05) & 27.24$\pm$0.59(10.84) \\
   \nodata  &  90.0\% & 26.94$\pm$0.80(11.05) & 21.63$\pm$0.56( 8.12) & 26.32$\pm$0.75(10.29) & 33.00$\pm$1.14(12.70) \\
   \nodata  & 100.0\% & 39.57$\pm$3.63(12.91) & 33.75$\pm$3.30( 9.75) & 38.98$\pm$3.61(12.17) & 45.99$\pm$3.98(14.82) \\
\enddata
\end{deluxetable}

\begin{deluxetable}{llccc}
\tabletypesize{\small}
\tablewidth{14cm}
\tablecaption{Fraction of galaxy type in each luminosity range shown in Figure \ref{Iint_I26_Itot}. 
\label{Type_lum}}
\tablehead{Hubble & \# of & 
\multicolumn{3}{c}{Fraction of galaxies within each ${\rm M_B}$~bin\tablenotemark{a} } \\
Type & galaxies &  $-18.12<{\rm M_B}$ & $-20.22<{\rm M_B}\le-18.12$ & ${\rm M_B}\le-20.22$ \\}
\startdata
S0-Sab  & 18 & 0.50 & 0.39 & 0.11 \\
Sb-Sd  & 61 &  0.26 & 0.48 & 0.26 \\
Sdm-Im-Pec & 22 & 0.82 & 0.18 & 0.00\\
\enddata
\tablenotetext{a}{We assume M$_{*}=-20.22$ 
\citep[][; after conversion to our adopted cosmology]{Marzke98}.}
\end{deluxetable}


\begin{thebibliography}{}

\bibitem[Aller(1942)]{Aller42} Aller, L. H. 1942, \apj, 95, 52

\bibitem[Baldry et al.(2002)]{Baldry02} Baldry, I. K., et al. 2002, \apj, 
569, 582

\bibitem[Bendo et al.(2002)]{Bendo02} Bendo, G. J. et al. 2002, 
\aj, 124, 1380

\bibitem[Brinchmann et al.(2004)]{Brinchmann04} Brinchmann, J., Charlot, 
S., White, S. D. M., Tremonti, C., Kauffmann, G., Heckman, T., \& 
Brinkmann, J. 2004, \mnras, 351, 1151

\bibitem[Burnstein \& Heiles(1984)]{Burnstein84} Burnstein, D. \& Heiles, C. 
1984, \apjs, 54, 33

\bibitem[Bell et al.(2002)]{Bell02} Bell, E. F., Gordon, K. D., Kennicutt, R. C., Jr., 
Zaritsky, D. 2002,  ApJ, 565, 994

\bibitem[Caplan \& Deharveng(1986)]{Caplan86} Caplan, J. \& Deharveng, L. 
1986, \aap, 155, 297

\bibitem[Cardelli, Clayton, \&Mathis(1989)]{Cardelli89} Cardelli, J. A., 
Clayton, G. C., \& Mathis, J. S. 1989, \apj, 345, 245 

\bibitem[Churches, Nelson, \& Edmunds(2001)]{Churches01} Churches, D. K., 
Nelson, A. H., \& Edmunds, M. G. 2001, \mnras, 327, 610

\bibitem[Consid\`{e}re et al.(2000)]{Considere00} Consid\`{e}re, S., 
Coziol, R., Contini, T., \& Davoust, E. 2000, \aap, 356, 89

\bibitem[Davis \& Peebles(1983)]{Davis83} Davis, M. \& Peebles, P. J. M.
1983, \apj, 267, 465

\bibitem[Dopita et al.(2000)]{Dopita00}  Dopita, M. A., Kewley, L. J.,
Heisler, C. A., \& Sutherland, R. S. 2000, \apj, 542,224

\bibitem[Edmunds(1984)]{Edmunds84} Edmunds, B. E. J. 1984, \mnras, 
211, 507

\bibitem[Edmunds \& Greenhow(1995)]{Edmunds95} Edmunds, M . G. \& Greenhow, R. M. 1995, \mnras, 272, 241

\bibitem[Garnett \& Shields(1987)]{Garnett87} Garnett, D. R. \& Shields,
. A. 1987, \apj, 317, 82

\bibitem[G\'{o}mez et al.(2003)]{Gomez03} G\'{o}mez, P. L. et al. 2003, 
\apj, 584, 210

\bibitem[Guesten \& Mezger(1982)]{Guesten82} Guesten, R. \& Mezger, P. G. 
1982, Vistas Astron., 26, 159

\bibitem[Hameed \& Devereux(1999)]{Hameed99} Hameed, S. \& Devereux, N. 
1999, \aj, 118, 730

\bibitem[Henry \& Worthey(1999)]{Henry99} Henry, R. B. C \& Worthey, G. 
1999, \pasp, 111, 919

\bibitem[Hill et al.(1999)]{Hill99} Hill, T. L., Heisler, C. A., Sutherland,
R. S., \& Hunstead, R. W. 1999, \aj, 117, 111

\bibitem[Hodge(1963)]{Hodge63} Hodge, P. W. 1963, \aj, 68, 237

\bibitem[Hopkins et al.(2003)]{Hopkins03} Hopkins, A. M., Miller, C. J., 
Nichol, R. C., Connolly, A. J., Bernardi, M., G\'{o}mez, P. L., Goto, T., 
Tremonti, C. A., Brinkmann, J., Ivezic, Z., \& Lamb, D. Q. 2003, \apj, 
599, 971

\bibitem[Huchra et al.(1983)]{Huchra83} Huchra, J., Davis, M., Latham, D., 
\& Tonry, J. 1983, \apjs, 52, 89

\bibitem[Israel \& Kennicutt(1980)]{Israel80} Israel, F. P. \& Kennicutt,
R. C. 1980, \apjl, 21, 1

\bibitem[James et al.(2004)]{James04} James, P. A. et al. 2004, \aap, 414, 23

\bibitem[Jansen et al.(1994)]{Jansen94} Jansen, R. A., Knapen, J. H., Beckman, J. E., Peletier, R. F. \& Hes, R. 1994, \mnras, 270, 373

\bibitem[Jansen et al.(2000a)]{Jansen00a} Jansen, R. A., Franx, M.,
Fabricant, D., \& Caldwell, N. 2000a, \apjs, 126, 271

\bibitem[Jansen et al.(2000b)]{Jansen00b} Jansen, R. A., Fabricant, D.,
 Franx, M., \& Caldwell, N. 2000b, \apjs, 126, 331

\bibitem[Jensen, Strom \& Strom(1976)]{Jensen96} Jensen, E. B., Strom, 
K. M., \& Strom, S. E. 1976, \apj, 209, 748

\bibitem[Kaufman et al. (1987)]{Kaufman87} Kaufman, M., Bash, F., 
Kennicutt, R. C., \& Hodge, P. W. 1987, \apj, 319, 61

\bibitem[Kennicutt(1998)]{Kennicutt98} Kennicutt, R. C. Jr. 1998, \araa,
36, 189

\bibitem[Kennicutt, Keel \& Blaha(1989)]{Kennicutt89} Kennicutt, R. C., 
Keel, W. C., \& Blaha, C. A. 1989, \aj, 97, 1022

\bibitem[Kewley et al.(2002)]{Kewley02a} Kewley, L. J., Geller, M. J., 
Jansen, R. A., \& Dopita, M. A. 2002 \aj, 124, 3135

\bibitem[Kewley \& Dopita(2002)]{Kewley02b} Kewley, L. J., \& 
Dopita, M. A., 2002 \apjs, 142, 35


\bibitem[Kewley et al.(2001a)]{Kewley01a} Kewley, L. J., Dopita, M. A.,
Sutherland, R. S., Heisler, C. A., \& Trevena, J. 2001a, \apj, 556, 121

\bibitem[Kewley et al.(2001b)]{Kewley01b} Kewley, L. J., Heisler, C. A.,
Dopita, M. A., \& Lumsden, S. 2001b, \apjs, 132, 37

\bibitem[Kobulnicky \& Kewley(2004)]{Kobulnicky04} Kobulnicky, H. A. \& Kewley, L. J.  2004,
ApJ, in press, astro-ph/0408128

\bibitem[Lamareille et al.(2004)]{Lamareille04} Lamareille, F., Mouhcine, M., Continu, T., Lewis, I., \& Maddox, S. 2004, astro-ph/0401615

\bibitem[Lehnert \& Heckman(1996)]{Lehnert96} Lehnert, M. D. \& Heckman, 
T. M. 1996, \apj, 462, 651

\bibitem[Maier, Meisenheimer, \& Hippelein(2004)]{Maier04} Maier, C., 
Meisenheimer, K., \& Hippelein, H. 2004, astro-ph/0401619

\bibitem[Martin \& Roy(1994)]{Martin94} Martin, P. \& Roy, J.-R. 1994, \apj, 424, 599

\bibitem[Marzke, Huchra, \& Geller(1994)]{Marzke94} Marzke, R. O., 
Huchra, J. P., \& Geller, M. J. 1994, \apj, 428, 43

\bibitem[Marzke et al.(1998)]{Marzke98} Marzke, R. O.,  da Costa, L. N., 
 Pellegrini, P. S., Willmer, C. N. A., Geller, M. J. 1998, \apj, 503, 617

\bibitem[Matteucci \& Fran\c{c}ois(1989)]{Matteucci89} Matteucci, F., Fran\c{c}ois, P. 1989, \mnras, 239, 885

\bibitem[Moll\'{a}, Ferrini, \& D\'{\i}az(1996)]{Molla96} Moll\'{a}, M., 
Ferrini, F., \& D\'{\i}az, A, I. 1996, \apj, 466, 668

\bibitem[M\"{o}llenhoff(2004)]{Mollenhoff04} M\"{o}llenhoff, C. 2004, 
\aap, 415, 63

\bibitem[Nakamura et al.(2003)]{Nakamura03} Nakamura, O., Fukugita, M., 
Brinkmann, J., \& Schneider, D. P. 2003, astro-ph/0312519

\bibitem[Oey \& Kennicutt(1993)]{Oey93} Oey, M. S. \& Kennicutt, R. C. 1993, 
\apj, 411, 137

\bibitem[Osterbrock(1989)]{Osterbrock89} Osterbrock, D. E. 1989, Astro
physics of Gaseous Nebulae and Active Galactic Nuclei (Mill Valley; University
Science Books)

\bibitem[Pagel et al.(1979)]{Pagel79} Pagel, B. E. J., Edmunds, M. G., Blackwell, D. E., Chun, M. S., \& Smith, G. 1979, \mnras, 189, 95

\bibitem[Pagel(1989)]{Pagel89} Pagel, B. E. J. 1989, RMxAA, 18, 161

\bibitem[P\'{e}rez-Gonz\'{a}lez et al.(2003)]{Perez03} 
P\'{e}rez-Gonz\'{a}lez, P. G., Zamorano, J., Gallego, J., 
Arag\'{o}n-Salamanca, A., Gil de Paz, A. 2003, \apj, 591, 827

\bibitem[Phillipps \& Edmunds(1991)]{Phillipps91} Phillipps, S. \& Edmunds, 
M. G. 1991, \mnras, 251, 84

\bibitem[Roberts, Huntley, \& van Albada(1979)]{Roberts79} Roberts, W. W., 
Huntley, J. M., \& van Albada, G. D. 1979, \apj, 233, 67

\bibitem[Roy\& Walsh(1997)]{Roy97} Roy, J. R. \& Walsh, J. R. 1997, 
\mnras, 288, 715

\bibitem[Sarazin(1976)]{Sarazin76} Sarazin, C. L. 1976, \apj, 208, 323

\bibitem[Schlegel, Finkbeiner \& Davis(1998)]{Schlegel98} Schlegel, D. J., 
Finkbeiner, D. P., \& Davis, M. 1998, \apj, 500, 525

\bibitem[Schulte-Ladbeck et al.(2004)]{Schulte04} Schulte-Ladbeck, R. E., 
Miller, C. J., Hopp, U., Hopkins, A., Nichol, R. C., Voges, W., \&
Taotao, F. 2004, astro-ph/0312069

\bibitem[Searle(1971)]{Searle71} Searle, L. 1971, \apj, 168, 327

\bibitem[Shields(1990)]{Shields90} Shields, G. A. 1990, \araa, 28, 525

\bibitem[Smith(1975)]{Smith75} Smith, H. E. 1975, \apj, 199, 591

\bibitem[Spergel et al.(2003)]{Spergel03} Spergel, D. N., Verde, L., Peiris, 
H. V., et al. 2003, astro-ph/030209 v3

\bibitem[Storchi-Bergmann(1991)]{Storchi91} Storchi-Bergmann, T. 1991,  \mnras, 
249, 404

\bibitem[Tinsley(1971)]{Tinsley71} Tinsley, B. M. 1971, \apss, 12, 394

\bibitem[Tremonti et al.(2004)]{Tremonti04} Tremonti, C. A., 
Heckman, T. M., Kauffmann, G., Brinchmann, J., Charlot, S., White, S. D. M.,  Seibert, M., Peng, E. W., Schegel, D. J., Uomoto, A., Fukugita, M., \&
Brinkmann, J. 2004, \apj, {\it submitted}

\bibitem[van der Hulst et al.(1988)]{vanderHulst88} van der Hulst, J. M., 
Kennicutt, R. C., Crane, P. C., \& Rots, A. C. 1988, \aap, 195, 38

\bibitem[de Vaucouleurs(1961)]{deVaucouleurs61} de Vaucouleurs, G. 1961, 
\apjs, 5, 233

\bibitem[de Vaucouleurs et al.(1991)]{deVaucouleurs91}de Vaucouleurs, G., 
de Vaucouleurs, A., Corwin, H. G., Jr., Buta, R. J.,
    Paturel, G., \& Fouqu\'{e}, P. 1991, Third Reference Catalogue of Bright
    Galaxies (New York: Springer)

\bibitem[Viallefond \& Goss(1986)]{Viallefond86} Viallefond, F. \& Goss, 
W. M. 1986, \aap, 154, 357

\bibitem[V\'{\i}lchez et al.(1988)]{Vilchez88} V\'{\i}lchez, J. M., Pagel,
B. E. J., Diaz, A. I., Terlevich, E., \& Edmunds, M. G. 1988, \mnras, 
235, 633

\bibitem[Zaritsky, Zabludoff, \& Willick(1995)]{Zaritsky95} Zaritsky, D., 
Zabludoff, A. I., \& Willick, J. A. 1995, \aj, 110, 1602

\end{thebibliography}
\end{document}